\documentclass[fleqn,usenatbib]{mnras}

\usepackage[T1]{fontenc}
\usepackage{ae,aecompl}

\usepackage{amsmath}
\usepackage{amssymb}

\usepackage{graphicx}

\usepackage{upgreek}
\renewcommand{\pi}{\uppi} 
\def\equationautorefname~#1\null{Equation~(#1)\null}

\usepackage{hhline}

\newcommand{\tauP}{\tau_{\mathrm{P}}} 
\newcommand{\Pfp}{P_{\mathrm{fp}}}
\newcommand{\Pfpdot}{\dot{P}_{\mathrm{fp}}}
\newcommand{\tauAge}{\tau_{\mathrm{age}}}
\newcommand{\tauEpsilon}{\tau_{\epsilon}}
\newcommand{\tauTheta}{\tau_{\theta}}

\newcommand{\tref}{t_{\textrm{ref}}}

\newcommand{\epsilonP}{\epsilon_\mathrm{p}}
\newcommand{\epsilondotP}{\dot{\epsilon}_\mathrm{p}}

\newcommand{\Tobs}{T_{\mathrm{obs}}}

\newcommand{\nudot}{\dot{\nu}}
\newcommand{\nuo}{\nu_{0}}
\newcommand{\nudo}{\dot{\nu}_{0}}
\newcommand{\nuddo}{\ddot{\nu}_{0}}

\newcommand{\M}{\mathcal{M}}
\newcommand{\data}{\mathcal{D}}
\newcommand{\params}{\boldsymbol{\lambda}}

\newcommand{\modelEpsDot}{$\epsilondotP$}
\newcommand{\modelThetaDot}{$\dot{\theta}$}
\newcommand{\modelThetaEpsDot}{$\{\dot{\theta},\epsilondotP\}$}
\newcommand{\modelEpsJumps}{$\Delta\epsilonP$}

\newcommand{\secs}{\mathrm{s}}
\newcommand{\years}{\mathrm{years}}

\def\LSWindowSizeMean{2058}
\def\LSWindowSizeStd{31}

\def\LSTspan{3200}
\def\Pfirst{505}
\def\Plast{470}

\def\ThetaDotPtauThetaMedianYears{307.7}

\def\ThetaDottauThetaFiveYears{114.3}
\def\EpsilonDottauEpsilonYearsMedian{213}
\def\EpsilonDottauEpsilonYearsStd{10}

\def\EpsilonDotThetaDotPtauThetaMedianYears{450.2}
\def\EpsilonDotThetaDottauEpsilonMedianYears{213}
\def\EpsilonDotThetaDottauEpsilonStdYears{10}

\def\EpsilonDotThetaDottauThetaFiveYears{170.9}
\def\oddsEpsilonDotBaseModel{73.65}
\def\errEpsilonDotBaseModel{0.97}

\def\oddsThetaDotBaseModel{-1.70}
\def\errThetaDotBaseModel{1.39}
\def\oddsEpsilonDotThetaDotBaseModel{72.45}
\def\errEpsilonDotThetaDotBaseModel{0.96}
\def\oddsEpsilonDeltasixJumpsBaseModel{73.53}
\def\errEpsilonDeltasixJumpsBaseModel{2.79}
\def\oddsEpsilonDotEpsilonDeltasixJumps{0.11}
\def\errEpsilonDotEpsilonDeltasixJumps{2.87}

\def\oddsEpsilonDotSpindownOnlyBaseModelSpindownOnly{49.35}
\def\errEpsilonDotSpindownOnlyBaseModelSpindownOnly{1.44}
\def\oddsEpsilonDotBeamwidthOnlyBaseModelBeamwidthOnly{23.46}
\def\errEpsilonDotBeamwidthOnlyBaseModelBeamwidthOnly{1.83}

\defcitealias{ashton2016}{Paper~I}

\newcommand{\selectedposteriorscaption}[2]{
Posterior probability distributions for the #1 in the #2-model.}
\newcommand{\completepriorposteriorcaption}[1]{
Prior distributions and a posterior distribution summary
for the #1-model parameters.}

\title[On the free-precession candidate PSR B1828-11]
      {On the free-precession candidate PSR B1828-11: Evidence for increasing
       deformation}
\author[G. Ashton, D. I. Jones and R. Prix]{
G.~Ashton,$^{1, 2}$\thanks{E-mail: gregory.ashton@ligo.org}
D.I.~Jones,$^{1}$
R.~Prix$^{2}$
\\
$^{1}${Mathematical Sciences and STAG Research Centre,
       University of Southampton,
       Southampton SO17 1BJ} \\
$^{2}${Max Planck Institut f{\"u}r Gravitationsphysik
       (Albert Einstein Institut) and Leibniz Universit\"at Hannover,
       30161 Hannover, Germany}
}

\pubyear{2016}

\begin{document}
\label{firstpage}
\pagerange{\pageref{firstpage}--\pageref{lastpage}}
\maketitle

\begin{abstract}

We observe that the periodic variations in spin-down rate and beam-width of the
radio pulsar PSR~B1828-11 are getting faster.  In the context of a free
precession model, this corresponds to a decrease in the precession period
$\Pfp$.    We investigate how a precession model can account for such a
decrease in $\Pfp$, in terms of an \emph{increase} over time in the absolute
biaxial deformation ($|\epsilonP|{\sim}10^{-8}$) of this pulsar.  We perform a
Bayesian model comparison against the `base' precession model (with constant
$\epsilonP$) developed in \cite{ashton2016}, and we obtain decisive odds in
favour of a time-varying deformation.  We study two types of time-variation:
(i) a linear drift with a posterior estimate of
$\epsilondotP{\sim}10^{-18}\,\mathrm{s}^{-1}$ and odds of $10^{75}$ compared to
the base-model, and (ii) $N$ discrete positive jumps in $\epsilonP$ with very
similar odds to the linear $\epsilonP$-drift model. The physical mechanism
explaining this behaviour is unclear, but the observation could provide a
crucial probe of the interior physics of neutron stars.  We also place an upper
bound on the rate at which the precessional  motion is damped, and translate
this into a bound on a dissipative mutual friction-type coupling between the
star's crust and core.

\end{abstract}


\section{Introduction}
\label{sec: introduction}

The ${\sim}500$~day periodicity observed in the timing properties and pulse
profile of PSR~B1828-11 provides a unique opportunity to test neutron star
physics.  The first model, proposed by \citet{bailes1993}, consisted of a
system of planets orbiting the pulsar. This model later lost favour, after
\citet{stairs2000} observed correlated modulation in the timing properties and
beam-shape (the ratio of the heights of two fitted integrated pulse
profiles). As such, a planetary model would require at least two orbiting
planets with orbital frequencies that differ by a factor of 2 (see for
example \citet{beauge2003extrasolar}), while both interact with the
magnetosphere over distances comparable to the Earth's orbit.

Instead, \citet{stairs2000} proposed that the star was undergoing free
precession, corresponding to a star that is deformed, with its  spin-vector and
angular momentum vectors misaligned.   Subsequent modelling by
\citet{jones2001}, \citet{link2001} and \citet{akgun2006} refined the
precessional  description, examining how the precessional motion served to
amplify the modulations in spin-down rate, providing some quantitative detail
to the precessional interpretation.

The existence of long period free precession has implications for the
interaction between the superfluid, superconducting and `normal' parts of the
star.  As shown by \citet{shaham1977}, a pinned superfluid, as typically
invoked to explain pulsar glitches, would result in a rather short free
precession period, so that the observed long period can be used to place upper
limits on the amount of pinned vorticity in PSR B1828-11; see
\citet{jones2001}, \citet{link2001} and \cite{linkcutler2002}.  Furthermore,
the interaction between neutron vortices and magnetic flux tubes in the stellar
core is likely to be highly dissipative, which led to \citet{link2003} drawing
the interesting conclusion  that the persistence of the free precession
required that neutron superfluidity and proton type II superconductivity
coexist nowhere in the star, or else that the superconductivity is of type I.
Additionally, \citet{wasserman_2003} has argued that a sufficiently strong
magnetic deformation of the stellar structure might force the star to undergo
free precession.
The issue of whether or not PSR B1828-11 really is precessing is therefore very
important, in terms of its microphysical implications.

Motivated by the existence of periodic nulling pulsars (such as PSR~B1931+24
\citep{kramer2006}), \citet{lyne2010} posited an alternative explanation for the modulations seen in PSR~B1828-11.  Namely, that the system is undergoing \emph{magnetospheric
switching}. In this model, the magnetosphere abruptly \emph{`state changes on a
fast time scale, but can then be stable for many months or years before
undergoing another fast change'} \citep{lyne2010}. This cycle periodically
repeats according to some clock and produces correlated changes in the timing
properties and pulse profile due to changes in the electromagnetic torque and
flow of charged particles.  However, to explain the double-peaked spin-down
rate of PSR~B1828-11, the model requires a complicated switching pattern such
as that proposed by \citet{perera2015}.



In addition to the long timescale modulations, PSR~B1828-11 is also known to
undergo short timescale (over periods of a few hours) switching in its
beam-shape, first demonstrated in \citet{stairs2003}, and illustrated further by  \citet{lyne2013}.
In the context of magnetospheric switching, the natural explanation is that, rather than remaining
in a single state for a prolonged period of time, the magnetosphere undergoes a
random process of flickering between two states.   

However,  the magnetospheric switching model does not provide an explanation of why the
modulations should be quasi-periodic.  To remedy this, \citet{jones2012}
proposed a model in which magnetospheric switching did indeed take place, but
precession provided the necessary clock mechanism, with the energies available
to accelerate particles in the magnetosphere being a function of the
precessional phase.  If there exists some critical energy threshold in the
magnetosphere, the precession model could then lead to sharp magnetospheric
transitions, with the magnetosphere being more likely to be in a given state at some precessional phases than others.   More generally, \citet{cordes2013} has argued that a component of pulsar timing noise can be attributed to pulsars making random transitions between two or more states, with a periodic bias active in some,  producing the observed quasi-periodicities.

It should also be noted that \citet{akgun2006} have argued that  short timescale
variations do not preclude the pure precession model (i.e. precession without any
magnetospheric switching) as a patchy emission region can also produce  short term variations in the beam-shape.

In an attempt to shed further light on the problem, in \citet{ashton2016} (hereafter
referred to as \citetalias{ashton2016}) we performed a Bayesian model
comparison using the \citet{lyne2010} spin-down rate and beam-width data
($W_{10}$, the width of the pulse at 10\% of the maximum) for PSR~B1828-11. We
compared a switching model to a precession model (neglecting the short term
flickering data and focusing only on the long term evolution), and found odds
of $10^{2.7\pm0.5}$ (`modest evidence') in favour of the precession model.

In this paper we will study what further inferences can be made based on simple some
generalisations of the precession model. We use the same data set (spanning
5280~days between MJD~49710 and MJD~54980) as in \citetalias{ashton2016}, which
was kindly provided by Andrew Lyne and originally published in
\citet{lyne2010}.  Specifically, we will look to see if there is any evidence
for time evolution in the amplitude of the precession, as measured by the
`wobble angle'  (see \autoref{sec: base-model} below), or for evolution in
the modulation period of the variations in spin-down and beam-width.  That the
amplitude of the precession might evolve is natural, as one would expect
dissipative processes within the star to damp the precession
\citep{sedrakian1999}.  That the modulation period might change is less
natural, but, as we describe in \autoref{sec: agnostic}, the data clearly
favour such an interpretation, so this needs to be included on the model.

The structure of this paper is as follows.  In \autoref{sec: agnostic} we
provide a model-independent demonstration that the modulation period of the
spin-down rate of PSR~B1828-11 is decreasing.  In \autoref{sec:
methodology} we describe our Bayesian methodology.  In \autoref{sec:
base-model} we describe our `base model' that other models will be compared to.
In \autoref{sec: evolution of the wobble angle} and \ref{sec: continuous
evolution of epsilon} we describe extensions of our base model where the wobble
angle and deformation, respectively, are allowed to vary (linearly) in time,
while in \autoref{sec: evolution of the wobble angle and deformation} we
allow both to parameters to vary.  In \autoref{sec: resolved discrete jumps
in epsilon} we consider a model where the deformation evolves by a series of
discrete jumps, rather than varying continuously.  In \autoref{sec:
theta_dot_interpretation} and \ref{sec: interpretation} we provide some
astrophysical interpretation of our results, and conclude in \autoref{sec:
discussion} with some discussion of implications of our work, and other
possible lines of attack.

In a separate paper \citep{letter} we discuss consistency requirements between
the free precession model of PSR~B1828-11 explored here and the glitch that
this pulsar underwent in 2009 (\citet{espinoza2011} and
\url{www.jb.man.ac.uk/~pulsar/glitches/gTable.html}).

\section{Model-independent evidence for a decreasing modulation period}
\label{sec: agnostic}
The modulation period of PSR~B1828-11 has so far been assumed constant.
However, we now show in a model-independent way that the period of the
spin-down-rate modulations in PSR~B1828-11 is getting shorter.

Let us define $\Delta\dot{\nu}$ as the \emph{spin-down rate residual}: the
result of removing a first-order polynomial from the spin-down rate (which can
be seen in Figure~1 of \citetalias{ashton2016}). This
discards information on the average spin-down rate and the second-order
spin-down rate $\ddot{\nu}$ leaving only the periodic modulations. To calculate
the period of modulations, we will apply a Lomb-Scargle periodogram to the
spin-down rate residual, which estimates the spectrum of periods by a
least-squares fit of sinusoids (in particular, we use the \textsc{scipy} \citep{scipy}
implementation of the \citet{townsend2010fast} algorithm).  In \autoref{fig:
spectrum}A we show the resulting estimated spectrum for the entire data period,
which agrees with the equivalent result presented in the additional material of
\citet{lyne2010}. Two dominant modes are present in the spectrum: a major mode
at $\sim 500$~days and a minor mode at $\sim 250$~days.

\begin{figure}
\centering
\includegraphics{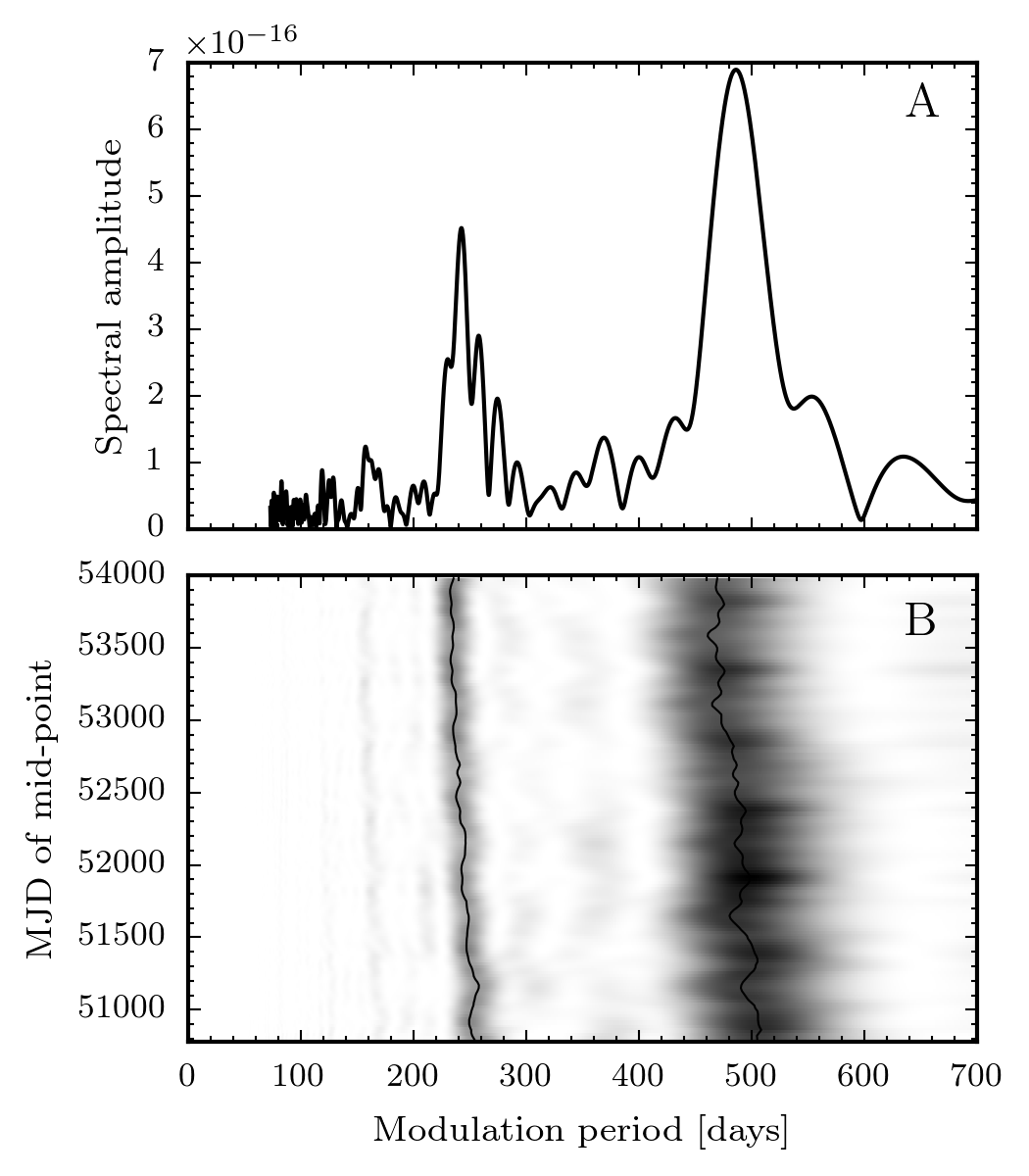}
\caption{A: The Lomb-Scargle estimate of the period spectrum of the
spin-down rate residual using the entire duration of data.
B: The period spectrum of the spin-down rate residual over a sliding window of
approximately \LSWindowSizeMean~days, as a function
of the window mid-point (on the y-axis).}
\label{fig: spectrum}
\end{figure}

To study how this spectrum varies with time, we apply the periodogram in a
sliding window across the spin-down rate residual data. Because the data is
unevenly sampled, it is not possible to use a fixed window size, but the
average window size is \LSWindowSizeMean ~days with a standard deviation of
\LSWindowSizeStd ~days.  This duration is sufficiently long to always include
several modulation cycles, but short enough to detect variations over the total
data span. To visualise the result, in \autoref{fig: spectrum}B we stack the
periodograms together and plot the spectral density as a function of the
mid-point of each time window.
This figure shows that the modulation period $P_{\rm mod}$ appears to be
\emph{decreasing} over time.  Taking the major mode from the first and last
sliding window we find that over a time span of $\LSTspan$~days the modulation
period decreased from $\Pfirst$ to $\Plast$~days, corresponding to a rate
of change of $\dot{P}_\textrm{mod} \approx -0.01$~s/s. We note that this
estimate is inherently imprecise due to the fact that the Lomb-Scargle method
is fitting a constant period sinusoid to data which is best described by a
sinusoid with changing period. Nevertheless, it does provide a rough estimate.
To underline the significance of this observed $\dot{P}_\textrm{mod}$, we found the
best-fit for a phenomological fixed-period sinusoidal model -- two sinusoids at
$P_\textrm{mod}$ and $P_{\textrm{mod}}/2$ with independent amplitudes and a
relative phase -- to the spin-down rate residual. We then generated $10^4$
realisations of central Gaussian noise with a standard deviation of
$4.3\times10^{-16}$~s$^{-2}$ (based on the standard-deviation of the residual
after removing the best-fit sinusoidal model). Adding the best-fit signal to
each noise realisation, we apply our Lomb-Scargle process to calculate the
change in period (due purely to the noise fluctuations) and find that the
maximum $|\dot{P}_\textrm{mod}| < 10^{-7}$. This illustrates that the
observed $\dot{P}_\textrm{mod}\sim-0.01$ for PSR~B1828-11 is highly unlikely to
be due to Gaussian noise fluctuations alone.

This shortening of the modulation period provides a new observational feature
that needs to be accommodated by any model trying to describe this data.  For
example in the planetary hypothesis this would require that the two planets
maintain orbital resonance while inspiralling.  For the magnetospheric
switching model proposed by \citet{perera2015} and further studied in
\citetalias{ashton2016}, it is unclear how this could be incorporated, given
the purely phenomenological nature of this model.  In the future it would
interesting to understand this observation in the context of other models;
in this work we explore how this feature is accommodated within the precession
model of \citetalias{ashton2016}.

\section{Data analysis methodology}
\label{sec: methodology}

In \citetalias{ashton2016} we performed a Bayesian model comparison between
precession (with non-circular beam geometry) and magnetospheric switching for
the observed long-term variations in spin-down rate and beam-width of
PSR~B1828-11.  Because of the purely phenomenological nature of the switching
model, no physical priors on its parameters were readily available and we
therefore resorted to a two-step approach: first we performed
parameter-estimation for both models on the spin-down data alone, by using wide
flat priors for both models.  Then we used the resulting posteriors as priors
for a model comparison on the beam-width data. This yielded odds of
$10^{2.7\pm0.5}$ in favour of the precession model.

In this work, we focus on physical generalisations of the precession model and
compare these to the `base' precession model.  The competing generalised
precession models share the parameters of the base-model, but extend them with
additional physical parameters that are allowed to be nonzero.  The base-model
priors can be thought of as effectively expressing certainty for these
additional parameters to vanish exactly, while the generalised models relax
this restriction and instead use plausible nonzero priors for them.  This
allows us to directly perform model comparison between base and generalised
models on the full data set comprising both spin-down and beam-width data.

We define the data $\data$ as $N$ observed $\nudot^{i}$ values and $M$ observed
$W_{10}^{j}$ values.  We denote as $\sigma_{\nudot}$ and $\sigma_{W_{10}}$ the
(assumed Gaussian) noise level for each type of observation.  The likelihood
for the data (see Section~2 of \citetalias{ashton2016}) given by model
$\mathcal{M}$ with model parameters $\params$ is then
\begin{align}
P(\data | \M, \vartheta) =
& {\prod_{i=1}^{N}}P(\dot{\nu}^{i}| \M, \params, \sigma_{\nudot})
{\prod_{j=1}^{M}}P(W_{10}^{j}| \M, \params, \sigma_{W_{10}}),
\end{align}
where $\vartheta = [\params, \sigma_{\nudot}, \sigma_{W_{10}}]$ is the full
set of parameters.
To approximate the posterior density of these parameters, we use the
\citet{foreman-mackay2013} implementation of the affine-invariant
parallel-tempered MCMC sampler \citep{goodman2010}; the exact methodology is
described in Appendix~A of \citetalias{ashton2016}. We then use
\emph{thermodynamic integration} \citep{goggans2004} to estimate the marginal
likelihood of a given model (see Section~4 of \citetalias{ashton2016}) and hence
the odds-ratio between models setting the prior ratio to unity. We use the
posterior odds between models to quantify how much, if at all, each extension
improves the power of the model to describe the data, compared to the
base-model.  This depends on both the improvement to fit the data as well as on
the respective prior volume of the extension parameters, which provides an
effective `Occam factor' against the extension.

\section{The precession base-model}
\label{sec: base-model}

We begin by introducing our \emph{base-model}, the precession model based on
the treatment given  in \citetalias{ashton2016}. It is against this which the
extended models will be compared.

\subsection{Defining the base-model}

We consider a biaxial star, spinning down by electromagnetic torque from the
magnetic dipole $\mathbf{m}$, which forms an angle $\chi$ with the symmetry
axis of the star. Following \citet{jones2001}, we define $\theta$ as the
\emph{wobble angle} between the symmetry axis and the angular momentum vector.
Precession produces modulations with period\footnote{In \citetalias{ashton2016}
we defined $\tauP$ as the precession period, here we will use $\Pfp$ in order
to be consistent with the literature.}$\Pfp$ in the rotation of the magnetic
axis. As a result, the spin-down rate and beam-width are modulated on the free
precession period.

Combining precession with a generalisation of the vacuum dipole torque and
allowing for an arbitrary braking index $n$, we show in Appendix~\ref{app:
derivation} that the spin-down rate, in the small-$\theta$ limit, is given by
\begin{align}
\begin{split}
\dot{\nu}(t) = & \nudo +  \nuddo (t - \tref) \\
 & - \nudo\theta \left[
2\cot\chi\sin(\psi(t)) - \frac{\theta}{2}\cos(2\psi(t))
\right],
\end{split}
\label{eqn: spin-down signal model}
\end{align}
where $[\dot{\nu}_0, \ddot{\nu}_0]$ are the fixed frequency derivatives defined
at a reference time $\tref$ and $\psi$ is one the three Euler angles describing
the orientation of the star (see for example \citet{landau1969}). We note that
\autoref{eqn: spin-down signal model} is equivalent to the results of
\citet{jones2001} and \citet{link2001}, although these previous works fixed the braking index to $n{=}3$.
If the spin-down age is much longer than the precession period $\Pfp$, we have
that
\begin{equation}
\psi(t)= -2\pi \, \frac{t-\tref}{\Pfp} + \psi_{0},
\label{eqn: psi(t)}
\end{equation}
in which we have implicitly defined the precession period as
\begin{align}
\Pfp = \frac{1}{\epsilonP \, \nu(t) \, \cos\theta},
\label{eqn: tauP definition}
\end{align}
where $\nu(t)$ is the instantaneous spin-frequency at time $t$, and
\begin{equation}
\label{eq:epsilon_P_definition}
\epsilonP = \frac{\Delta I_{\rm d}}{I_{\rm prec}} ,
\end{equation}
where $\Delta I_{\rm d}$ is the stellar deformation caused by elastic/magnetic
strains, while $I_{\rm prec}$ is that part of the star that participates in the
free precession.  We can expect $I_{\rm crust} < I_{\rm prec} < I_*$; see
\citet{jones2001} for details.

Formally, the spin frequency $\nu(t)$  is
the integral of \autoref{eqn: spin-down signal model}. However, the
sinusoidal variations due to precession will average to zero over an integer
number of cycles. Therefore, we will neglect the residual modulations, which
will have a negligible effect on the precession period, and approximate
the spin frequency in \autoref{eqn: tauP definition} by
\begin{align}
\nu(t) = \nuo + \nudo \, (t - \tref) + \frac{\nuddo}{2} \, (t-\tref)^{2},
\end{align}
where $\nuo$ is the fixed frequency of the star at $\tref$.  We will define
$\tref$ at the epoch given in the ATNF \citep{atnf} entry for PSR~B1828-11.
This reference time, the frequency and its derivatives, and other useful
quantities are listed in \autoref{tab: ATNF}.
\begin{table}
\centering
\caption{Table of astrophysical data for B1828-11 taken from the ATNF pulsar
catalogue \citep{atnf}, available at
\url{www.atnf.csiro.au/people/pulsar/psrcat}.}
\label{tab: ATNF}
\begin{tabular}{ll}
Parameter & ATNF value\\ \hline
$\tref$ & MJD 49621 \\
$\nuo$ & $2.46887171470 \pm 7 \times10^{-11}$ ~ Hz \\
$\nudo$ & $-3.658728\times10^{-13} \pm 5 \times10^{-19}$ ~ Hz/s\\
$\nuddo$ & $8.72\times10^{-25} \pm 9 \times10^{-27}$ ~ Hz/s$^{2}$\\
$\tauAge = -\nu_0/\dot{\nu}_0$ & $1.07\times10^{5}$~yrs\\
$n = \nuddo \nuo / \nudo^2$ & $16.08 \pm 0.7$ \\
Distance & 3.58 kpc
\end{tabular}
\end{table}

The pulse beam-width $W_{10}$ is defined as the width of the pulse at 10\% of
the observed peak intensity. This beam-width depends on the motion of the
dipole $\mathbf{m}$, how the intensity of emission varies across the beam, and
on the relative position of the observer and the beam.  The angle $\Theta$
between the dipole $\mathbf{m}$ and the angular momentum $\mathbf{J}$ can be
expressed as
\begin{align}
\Theta(t) = \cos^{-1}\left(\sin\theta\sin\chi\sin(\psi(t))
                           + \cos\theta\cos\chi\right),
\label{eqn: Theta}
\end{align}
which describes the polar motion of $\mathbf{m}$ in the  inertial
frame \citep{bisnovatyi1990model,jones2001}.  Let $\iota$ denote the angle
between the observing direction and $\mathbf{J}$, and so the latitudinal
separation between observer and beam is given simply by $\Delta\Theta(t) =
\Theta(t) - \iota$.

In \citetalias{ashton2016} we first considered an emission model where the
intensity of the emitted radiation is circularly symmetric around the dipole
$\mathbf{m}$ with a radial Gaussian fall-off.  However, this
simple model is unable to account for the observed variations in $W_{10}$, and
we therefore extended the model to allow for the longitudinal width of the Gaussian describing the intensity to depend
on the latitude $\Delta\Theta(t)$ of the cut made through the beam; this was
found to produce good agreement with observations (similar
conclusions have previously been obtained by \citet{link2001}).  This results
in a beam-width expression of the form
\begin{align}
W_{10}(t) = \frac{1}{\nu(t)\,\pi}\sqrt{\frac{2\ln10}{\sin\Theta(t) \,\sin\iota}}
 \left(\rho_2^{0} + \rho_2'' \, \Delta\Theta(t)^{2} \right),
\label{eqn: beam-width signal model}
\end{align}
where $\rho_2^0$ is the width of the Gaussian intensity at $\Delta\Theta=0$ and
$\rho_2''$ describes the variation in intensity with $\Delta\Theta$; see \citetalias{ashton2016}.  Our
formulation of the base-model is now complete: \autoref{eqn: spin-down signal
model} is the base spin-down model and \autoref{eqn: beam-width
signal model} is the base beam-width model.

This formulation of the base precession model differs from that used in
\citetalias{ashton2016} in two ways. First, in \citetalias{ashton2016},
$\Pfp$ was a constant model parameter. But in \autoref{eqn: tauP
definition}, we now express the precession period $\Pfp$ in terms of the
fundamental model parameters: the instantaneous spin-frequency $\nu(t)$, wobble
angle $\theta$, and the deformation $\epsilonP$. While this change of
parameterisation provides a more complete description (in that it includes the
time evolution of $\Pfp$ with $\nu(t)$), it was found to produce no significant
change in the fit. Second, the sign of the first term of \autoref{eqn:
psi(t)} was positive in Eqn.~(16) of \citetalias{ashton2016}, but is now
negative; this change amounts to a redefinition of $\Pfp$ which was done such
that for an oblate star, $\epsilonP$ and $\Pfp$ are both positive, while for a
prolate star both these quantities are formally negative. As the spin-down rate
and beam-width of the precession model (\autoref{eqn: spin-down signal
model} and \autoref{eqn: beam-width signal model} respectively) are
invariant to this change of sign (modulo addition of $\pi$ to $\psi_0$), the
redefinition of $\Pfp$ makes no substantial difference to the model.

The base-model and all extensions considered in this work are subject to
two symmetries which are important when interpreting our results.  First, as a
consequence of the invariant nature of the spin-down rate and beam-width to the
sign of $\epsilonP$, the data cannot fix the overall sign of $\epsilonP$.  We
restrict this symmetry by choosing $\epsilonP>0$ in the prior, but we note that
solutions where $\epsilonP \rightarrow -\epsilonP$ are equally valid. Second,
it was noted by \citet{arzamasskiy2015} that the spin-down rate in the
precession model is symmetric under the substitution $\theta \leftrightarrow
\chi$ (we discuss how this can be derived for \autoref{eqn: spin-down signal
model} in Appendix~\ref{app: derivation}); in our model, this is also true for
the beam-width. For both the spin-down and beam-width models, this is
fundamentally due to the symmetry of $\chi$ and
$\theta$ in \autoref{eqn: Theta}. In our analysis, we consider only the
`large $\chi$' model (as defined by \citet{arzamasskiy2015}) and restrict this
symmetry in the derivation by assuming that $\theta \ll 1$ and in the choice of
prior. But, rederiving the equations with $\chi \ll 1$ instead results in
\autoref{eqn: spin-down signal model} with $\theta \leftrightarrow \chi$.
Therefore, all models and parameter estimation considered in this
work can equally be applied to the `small $\chi$' model by interchanging $\chi$
and $\theta$. These symmetries may be important to consider when relating the
model extension to physical theories.

\subsection{Applying the base-model to the data}

The base-model consists of the spin-down and beam-width predictions given in
\autoref{eqn: spin-down signal model} and \autoref{eqn: beam-width signal
model}. Before applying these to the data, we first define our priors.
Since we will use the same priors for these parameters when considering
the extended models in the following sections, their prior volume won't have an
impact on the model-comparison odds.

The full set of priors are listed in \autoref{tab: base-model prior
posterior} and we now describe our choices in detail.  For the spin frequency
and frequency derivatives we apply astrophysical priors based on data from the
ATNF database (which is listed in \autoref{tab: ATNF}). Specifically, we use
normal distributions with mean and standard-deviation given by the ATNF values.
For the deformation $\epsilonP$ we use the absolute value of a normal
distribution as prior, ensuring our gauge choice of $\epsilonP\ge0$.  The normal distribution has zero mean, and a standard deviation of $10^{-8}$, the approximate known value of $\epsilonP$ \citepalias{ashton2016}.  For the angles $\theta$ and $\chi$
we restrict their domain to solutions where the wobble angle $\theta$ is small
while the magnetic inclination $\chi$ is close to orthogonal (the `large $\chi$'
model, for more details see Appendix~\ref{app: derivation})). The beam-width
parameters ($\rho_2^0$ and $\rho_2''$) use priors from \citetalias{ashton2016},
which were chosen to give a range of beam-widths up to $10\%$ of the period and
allow for some non-circularity.  Finally the phase is given a uniform prior
over its domain and we use uniform priors for $\sigma_{\dot{\nu}}$ and
$\sigma_{W_{10}}$ from a crude estimate of the data.
\begin{table}
\centering
\caption{\completepriorposteriorcaption{base}}
\label{tab: base-model prior posterior}
\scriptsize
\tabcolsep=0.07cm
\begin{tabular}{llll} \hhline{====}
        & Prior & Posterior median $\pm$ s.d. &  Units\\ \hline
$\nu_0$ & $\mathcal{N}$(2.46887171470, ${7.0}{\times} 10^{\textrm{-}11}$) & 2.47$\pm$${7.0}{\times} 10^{\textrm{-}11}$ & Hz \\
$\dot{\nu}_0$ & $\mathcal{N}$(${\text{\textrm{-}}3.658728}{\times} 10^{\textrm{-}13}$, ${5.0}{\times} 10^{\textrm{-}19}$) & ${\text{\textrm{-}}3.66}{\times} 10^{\textrm{-}13}$$\pm$${5.0}{\times} 10^{\textrm{-}19}$ & Hz/s \\
$\ddot{\nu}_0$ & $\mathcal{N}$(${8.72}{\times} 10^{\textrm{-}25}$, ${9.0}{\times} 10^{\textrm{-}27}$) & ${8.73}{\times} 10^{\textrm{-}25}$$\pm$${9.0}{\times} 10^{\textrm{-}27}$ & Hz/s$^{2}$ \\
$\epsilon_\mathrm{p}$
 & $|\mathcal{N}$(0, ${1}{\times} 10^{\textrm{-}8}$)$|$ & ${9.67}{\times} 10^{\textrm{-}9}$$\pm$${1.1}{\times} 10^{\textrm{-}11}$ &   \\
$\theta$
 & Unif(0, 0.1) & 0.0490$\pm$0.0020 & rad \\
$\chi$
 & Unif($2\pi/5$, $\pi/2$) & 1.5519$\pm$0.0013 & rad \\
$\psi_0$
 & Unif(0, $2\pi$) & 5.4821$\pm$0.0456 & rad \\
$\rho_{2}^{0}$
 & Unif(0, 0.1464) & 0.0246$\pm$0.0004 & rad \\
$\rho_{2}''$
 & $\mathcal{N}$(0, 6.83) & 3.36$\pm$0.4 & rad$^{\textrm{-}2}$ \\
$\cos(\iota)$
 & Unif(\textrm{-}1, 1) & ${7.57}{\times} 10^{\textrm{-}3}$$\pm$${2.1}{\times} 10^{\textrm{-}3}$ &   \\
$\sigma_{\dot{\nu}}$
 & Unif(0, ${1}{\times} 10^{\textrm{-}15}$) & ${4.09}{\times} 10^{\textrm{-}16}$$\pm$${1.9}{\times} 10^{\textrm{-}17}$ & $\mathrm{s}^{\textrm{-}2}$ \\
$\sigma_{W_{10}}$
 & Unif(0, ${5.0}{\times} 10^{\textrm{-}3}$) & ${1.59}{\times} 10^{\textrm{-}3}$$\pm$${4.3}{\times} 10^{\textrm{-}5}$ & s \\
\hhline{====}
\end{tabular}

\end{table}

\begin{figure}
\centering
\includegraphics[]{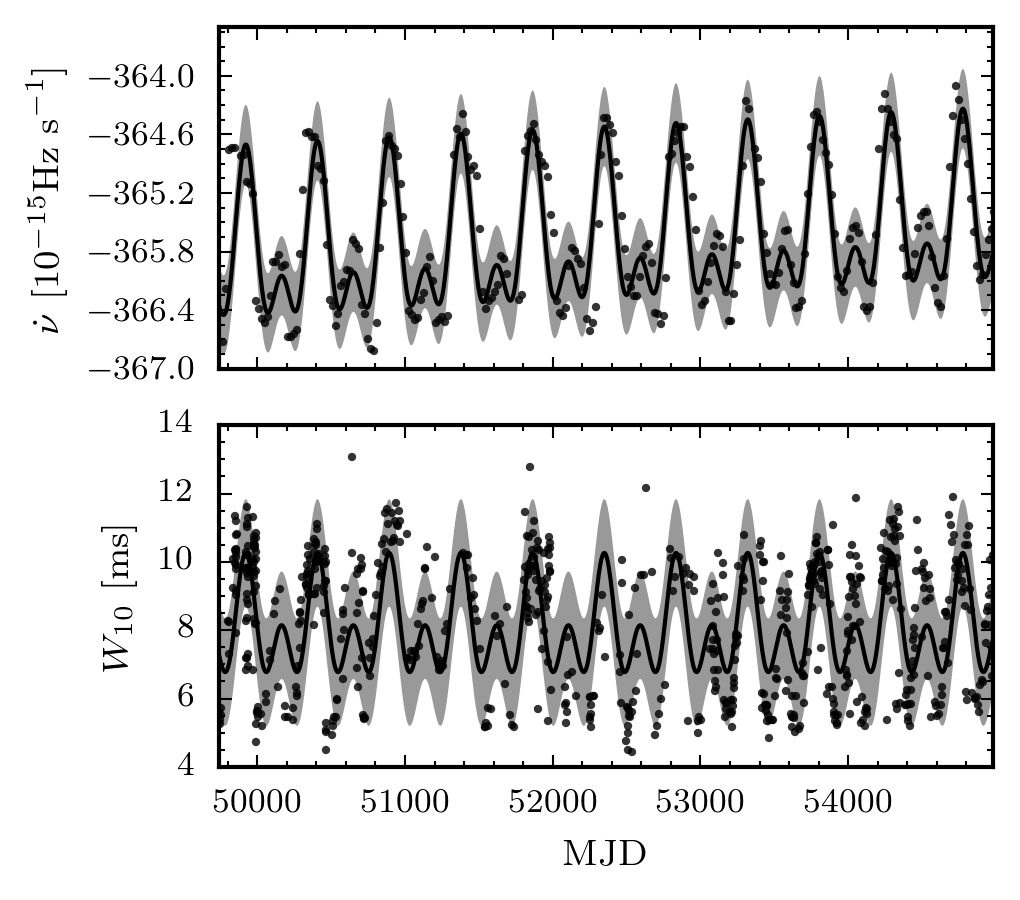}
\caption{Comparison between the base-model (solid line) using maximum-posterior
parameter estimates (MPE) and the observed spin-down and beam-width data (black
dots). The shaded region indicates $1\,\sigma$ of the estimated noise level in
the spin-down and beam-width data, respectively.}
\label{fig: base-model posterior fit}
\end{figure}

We run MCMC simulations applying the base-model to the data under these priors
and check that they converge and properly sample the posterior. In
\autoref{fig: base-model posterior fit} we show the spin-down and beam-width
data together with the \emph{maximum posterior estimate} (MPE) solution of the
model, i.e.\ using the parameters with the highest posterior probability.
The samples from the converged MCMC chains are used to estimate the posterior
distributions, which we find to be Gaussian-like, and which we summarise in
the second column of \autoref{tab: base-model prior posterior} by their
median and standard deviation.

Compared to \citetalias{ashton2016} this base-model already contains one
model extension: allowing for variation in $\Pfp$ due to $\nu(t)$, as seen in
\autoref{eqn: tauP definition}.  However, this does not make any appreciable
difference to the result in that there is no noticeable difference between the
two panels of \autoref{fig: base-model posterior fit} and Figure~7B and Figure~11B
of \citetalias{ashton2016}. Furthermore, this extension does not explain the
observed changing modulation period discussed in \autoref{sec: agnostic}.  In
order to see this quantitatively, we expand \autoref{eqn: tauP definition}
to first order as
\begin{align}
\Pfp \approx \frac{1}{\epsilonP \nuo \cos\theta}
              \left(1 - (t-\tref) \frac{\nudo}{\nuo}\right).
\end{align}
Since $\nudo < 0$ this produces an increasing precession period, which over the
observation span produces a fractional change in precession period of
$\sim7{\times}10^{-5}$. Hence, the effect of the spin-down is too small and of
the wrong sign to explain the observations of \autoref{sec: agnostic}.

From \autoref{eqn: tauP definition} we see that there are two further
possible ways that $\Pfp$ can evolve: either the wobble angle $\theta$ or the
deformation $\epsilonP$ must evolve (or both).  In the following sections, we
will consider these possibilities in turn and evaluate the improvements in the
power of the respective model to describe the data by computing their odds
compared to the base-model.

\section{Secular evolution of the wobble angle: the \modelThetaDot-model}
\label{sec: evolution of the wobble angle}

There are two reasons for allowing a secular evolution of the precession wobble
angle.  Firstly, from \autoref{eqn: tauP definition} we see that such an
evolution could potentially drive a change in the precession period explaining
the results of \autoref{sec: agnostic}. However, simple estimates show that
the required rate of variation in $\theta$ is much too large to be consistent
with the observations; we give such arguments in
\autoref{sect:ruling_out_theta} below.  Secondly, and perhaps more fundamentally,
in the precessional interpretation, dissipative processes are expected to exist
and should damp the wobble angle, which would provide insights into the
crust-core coupling (see for example \citet{sedrakian1999} and
\citet{levin2004}).

We model this in the simplest way by assuming that $\theta$ changes linearly in
time as
\begin{align}
\theta(t) = \theta + \dot{\theta} \, (t - \tref).
\label{eqn: theta extension}
\end{align}
The base-model spin-down rate of \autoref{eqn: spin-down signal model} was
derived under the assumption that $\theta$ is constant. However, when
rederiving this expression with an evolving $\theta$ according to
\autoref{eqn: theta extension}, we find that (to first order) the expression
remains valid with the simple substitution $\theta \rightarrow \theta(t)$.

\subsection{Can a changing $\theta$ explain the observed decrease in precession
            period?}
\label{sect:ruling_out_theta}

Using the following simple argument, we can see that a nonzero
$\dot{\theta}$ cannot consistently explain the observed decrease in precession
period of $\Pfpdot\approx-0.01$~s/s found in \autoref{sec: agnostic}.  Taking
the time-derivative of \autoref{eqn: tauP definition} with
$\theta=\theta(t)$ (and dropping a negligible contribution
$\Pfp/\tauAge\sim10^{-5}$~s/s to $\Pfpdot$) we can estimate the required
$\dot{\theta}$ as
\begin{equation}
  \dot{\theta}= \frac{\Pfpdot}{\Pfp}\,\cot\theta
  \approx -5\times10^{-9} \textrm{rad/s}\,,
\label{eqn: theta dot}
\end{equation}
therefore
\begin{equation}
  \tau_\theta \equiv \frac{\theta}{|\dot{\theta}|}
  = \frac{\theta}{\cot\theta}\,\frac{\Pfp}{|\Pfpdot|}
  \approx \frac{1}{400}\,\frac{\Pfp}{\Pfpdot}
  \sim0.34\,\years\,.
\label{eqn: tau theta requirement}
\end{equation}
where we used the base-model posterior estimates from \autoref{tab:
base-model prior posterior} for $\theta$ and for $\Pfp$ (these values are
derived assuming that $\dot{\theta} = 0$, however, as shown later in
\autoref{tab: theta dot prior posterior} they are consistent with those found
when this assumption is relaxed).  Similarly, with the estimate of
\autoref{eqn: theta dot} the predicted relative change in the spin-down
modulation amplitude from \autoref{eqn: spin-down signal model} over the
observation period of $T\approx5000$~days would amount to
\begin{align}
\frac{\dot{\theta}T}{\theta} \approx -46.8\,.
\end{align}
This level of change in $\theta$ is inconsistent with the observed spin-down
variations, which are well described by a model with an approximately constant
$\theta$ (e.g.\ see \autoref{fig: base-model posterior fit}).

We can therefore conclude that changes in $\theta$ are unable to explain the
decrease in modulation period.  Fundamentally, this stems from the fact that
the dependence of the modulation amplitude on $\theta$ is $\propto \theta$,
while the dependence of $\Pfp$ is~$\propto 1 + \theta^{2}/2$ for $\theta \ll
1$.

\subsection{Applying the \modelThetaDot-model to the data}

We choose a weakly-informative prior for the additional model parameter
$\dot{\theta}$: a central normal distribution with standard-deviation of
$2.2\times 10^{-10}$ rad/s, which is the value one would get if
$\dot{\theta}\sim 2\theta/T$, so effectively this allows $\theta$ to change by
twice its magnitude over the observation time $T$. Using such a wide prior
allows us to be confident that the posterior upper limit on $\dot{\theta}$
will be informed by the data and not the result of an overly-constrained prior.

\begin{figure}
\centering
\includegraphics[]{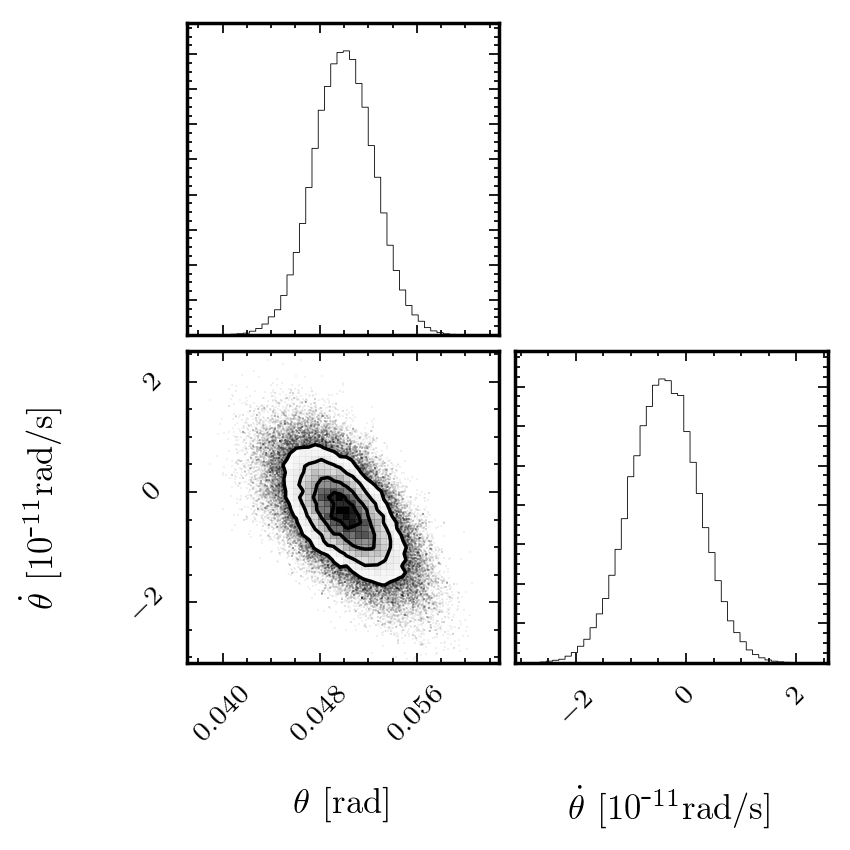}
\caption{\selectedposteriorscaption{wobble angle $\theta$ and its rate of
change $\dot{\theta}$}{\modelThetaDot}}
\label{fig: theta dot posterior}
\end{figure}

The resulting posteriors for $\theta$ and $\dot{\theta}$ are shown in
\autoref{fig: theta dot posterior} and the posteriors for all model parameters
are summarised in \autoref{tab: theta dot prior posterior} alongside the
priors (which are identical to those of the base-model).
\begin{table}
\centering
\caption{\completepriorposteriorcaption{\modelThetaDot}}
\label{tab: theta dot prior posterior}
\tabcolsep=0.07cm
\scriptsize
\begin{tabular}{llll} \hhline{====}
        & Prior & Posterior median $\pm$ s.d. &  Units\\ \hline
$\nu_0$ & $\mathcal{N}$(2.46887171470, ${7.0}{\times} 10^{\textrm{-}11}$) & 2.47$\pm$${7.0}{\times} 10^{\textrm{-}11}$ & Hz \\
$\dot{\nu}_0$ & $\mathcal{N}$(${\text{\textrm{-}}3.658728}{\times} 10^{\textrm{-}13}$, ${5.0}{\times} 10^{\textrm{-}19}$) & ${\text{\textrm{-}}3.66}{\times} 10^{\textrm{-}13}$$\pm$${5.0}{\times} 10^{\textrm{-}19}$ & Hz/s \\
$\ddot{\nu}_0$ & $\mathcal{N}$(${8.72}{\times} 10^{\textrm{-}25}$, ${9.0}{\times} 10^{\textrm{-}27}$) & ${8.73}{\times} 10^{\textrm{-}25}$$\pm$${8.9}{\times} 10^{\textrm{-}27}$ & Hz/s$^{2}$ \\
$\epsilon_\mathrm{p}$
 & $|\mathcal{N}$(0, ${1}{\times} 10^{\textrm{-}8}$)$|$ & ${9.67}{\times} 10^{\textrm{-}9}$$\pm$${1.2}{\times} 10^{\textrm{-}11}$ &   \\
$\theta$
 & Unif(0, 0.1) & 0.0500$\pm$0.0025 & rad \\
$\dot{\theta}$
 & $\mathcal{N}$(0, ${2.2}{\times} 10^{\textrm{-}10}$) & ${\text{\textrm{-}}3.97}{\times} 10^{\textrm{-}12}$$\pm$${6.3}{\times} 10^{\textrm{-}12}$ & rad/s \\
$\chi$
 & Unif($2\pi/5$, $\pi/2$) & 1.5519$\pm$0.0013 & rad \\
$\psi_0$
 & Unif(0, $2\pi$) & 5.4688$\pm$0.0494 & rad \\
$\rho_{2}^{0}$
 & Unif(0, 0.1464) & 0.0246$\pm$0.0004 & rad \\
$\rho_{2}''$
 & $\mathcal{N}$(0, 6.83) & 3.33$\pm$0.4 & rad$^{\textrm{-}2}$ \\
$\cos(\iota)$
 & Unif(\textrm{-}1, 1) & ${7.51}{\times} 10^{\textrm{-}3}$$\pm$${2.1}{\times} 10^{\textrm{-}3}$ &   \\
$\sigma_{\dot{\nu}}$
 & Unif(0, ${1}{\times} 10^{\textrm{-}15}$) & ${4.09}{\times} 10^{\textrm{-}16}$$\pm$${1.9}{\times} 10^{\textrm{-}17}$ & $\mathrm{s}^{\textrm{-}2}$ \\
$\sigma_{W_{10}}$
 & Unif(0, ${5.0}{\times} 10^{\textrm{-}3}$) & ${1.59}{\times} 10^{\textrm{-}3}$$\pm$${4.2}{\times} 10^{\textrm{-}5}$ & s \\
\hhline{====}
\end{tabular}

\end{table}

The $\theta$ posterior is found to be essentially unchanged with respect to the
base-model.  The $\dot{\theta}$ posterior shows a substantial amount of
`shrinkage' compared to its prior range, but is fully consistent with
$\dot{\theta}=0$ and therefore provides no evidence that $\theta$ is actually
changing.  Nevertheless, we can use this to place constraints on the timescale
of $\theta$-changes by defining $\tauTheta \equiv |\theta/\dot{\theta}|$ and
using the samples from \autoref{fig: theta dot posterior} to estimate the
posterior distribution for $\tauTheta$, which is shown in \autoref{fig: theta
dot tau Theta}.  This figure shows that there is little support for variation
timescales below $\sim100\,\years$ (confirming that the required timescale for
$\tauTheta$ to explain the changing modulation period given in \autoref{eqn:
tau theta requirement} is too short). The distribution has a long tail,
allowing for much longer timescales. The median of the distribution is
$\ThetaDotPtauThetaMedianYears\,\years$ and we can place a $95\%$ credible
lower limit of $\tauTheta > \ThetaDottauThetaFiveYears\,\years$.
\begin{figure}
\centering
\includegraphics[width=0.5\textwidth]{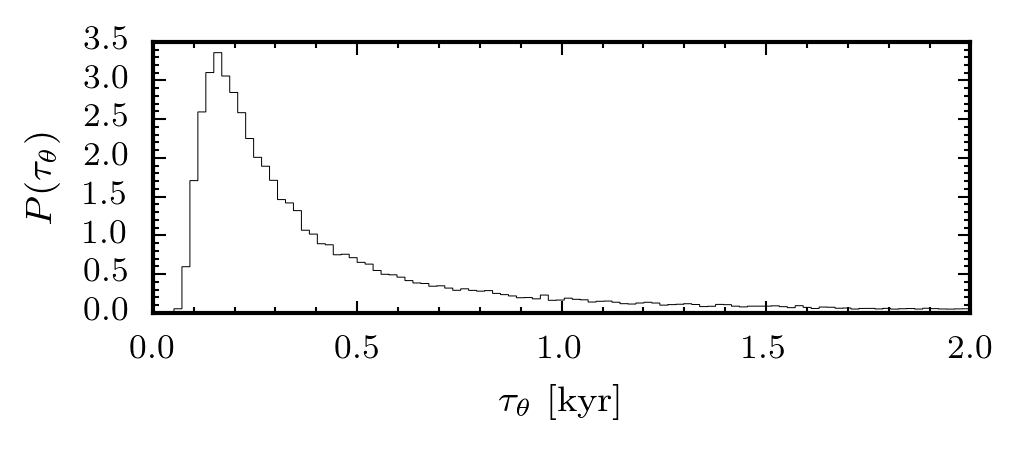}
\caption{The posterior distribution for $\tauTheta\equiv|\theta/\dot{\theta}|$
under the \modelThetaDot-model.}
\label{fig: theta dot tau Theta}
\end{figure}
The odds between the \modelThetaDot-model compared to the base-model are found
as $10^{\oddsThetaDotBaseModel \pm \errThetaDotBaseModel}$, i.e.\ weak evidence
against this extension.  This shows the effect of the built-in Bayesian `Occam
factor': the extension of allowing $\dot{\theta}\not=0$ (which can only
improve the fit to the data) does not provide a sufficient improvement in
likelihood compared to the increase in prior volume.

\section{Secular evolution of the deformation: the \modelEpsDot-model}
\label{sec: continuous evolution of epsilon}

After ruling out variations in $\nu$ and $\theta$ in the previous sections as
the cause for the observed level of $\Pfpdot$, we see from \autoref{eqn:
tauP definition} that this leaves only variations in the deformation
$\epsilonP$ as a possible explanation. In this section and \autoref{sec:
resolved discrete jumps in epsilon}, we consider two distinct types of
time-evolution in $\epsilonP$: firstly the \modelEpsDot-model, a slow
continuous change (approximated by the linear term) in $\epsilonP$, and then
the \modelEpsJumps-model, a series of distinct `jumps' in $\epsilonP$.  These
are just two possible phenomological models which are not founded in any
physical theory, instead they are chosen simply to model two distinctive
behaviours.

\subsection{Defining the \modelEpsDot-model}

We consider the simplest continuously changing deformation model by including a
linear term (which also describes a larger class of sufficiently-slow
continuous change in $\epsilonP$):
\begin{align}
\epsilonP(t) = \epsilonP + \epsilondotP \, (t - \tref).
\label{eqn: epsilon extension}
\end{align}
We will discuss some potential physical mechanisms for such a secular change in
\autoref{sec: interpretation}.

Allowing for a time-varying $\epsilonP(t)$ in \autoref{eqn: tauP definition}
and assuming this accounts for the majority of the change in $\Pfp$, we obtain
\begin{equation}
\label{eq:1}
\frac{\Pfpdot}{\Pfp} \approx -\frac{\epsilondotP}{\epsilonP} \equiv -\frac{1}{\tau_{\epsilon}}\,,
\end{equation}
where we have defined the characteristic timescale $\tau_{\epsilon}$ for the
rate of change in $\epsilonP$.

Given that $\Pfp$ is decreasing with time (c.f.\ \autoref{sec: agnostic}), for
$\epsilonP > 0$ this implies $\epsilondotP > 0$, while for $\epsilonP<0$ this
would correspond to $\epsilondotP < 0$.  As previously mentioned, we are unable
to determine the sign of $\epsilonP$ from our current precession model, but in
either case the magnitude of the deformation has to be increasing, i.e.\
$d|\epsilonP|/dt>0$, in order to account for the observed decrease in $\Pfp$.

From \autoref{eq:1} we can estimate the required $\epsilondotP$ for the
observed $\Pfpdot\approx-0.01$~s/s as found in \autoref{sec: agnostic}, which
yields $\epsilondotP\approx 2\times10^{-18}$~s$^{-1}$.  We use this as the
scale for a central Gaussian prior on $\epsilondotP$ as
\begin{align}
\epsilondotP \sim |N\left(0,\,2{\times}10^{-18}\right)|,
\label{eqn: epsilondot prior}
\end{align}
where we restrict ourselves to positive values in accordance to our gauge
choice of $\epsilonP>0$.

This prior is weakly informed by the data, but we could equally well consider a
less-informed choice of, say, allowing $\epsilonP$ to double in size over the
observation timescale $T=5000\,$days, which would yield a prior scale of
$\epsilondotP \sim 2{\times}10^{-17}\secs^{-1}$.  This is only a factor of 10
wider compared to \autoref{eqn: epsilondot prior}, and would be expected to
reduce the odds by about one order of magnitude at most via the larger `Occam
factor' (i.e.\ prior volume).  Re-running the analysis with the wider prior
confirms this, as we obtain odds that are reduced by a factor of $\sim5$
compared to using \autoref{eqn: epsilondot prior}, while yielding
essentially unchanged posteriors.

\subsection{Applying the \modelEpsDot-model to the data}

\begin{figure}
\centering
\includegraphics[]{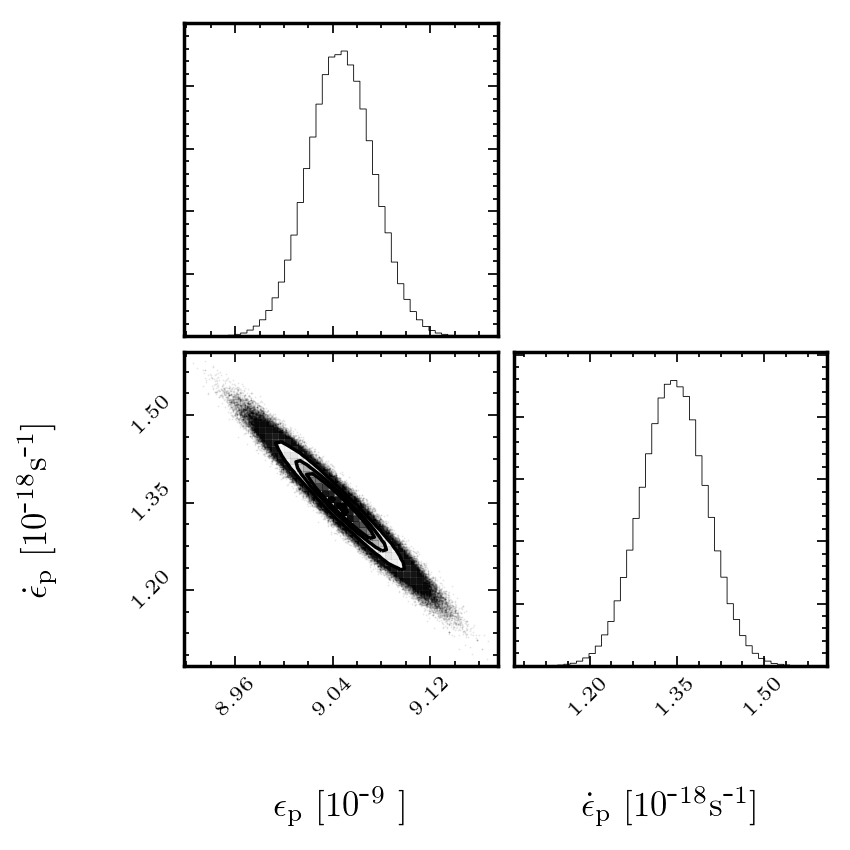}
\caption{\selectedposteriorscaption{deformation $\epsilonP$ and its rate of
change $\epsilondotP$}{\modelEpsDot}}
\label{fig: epsilon dot posterior}
\end{figure}

The estimated posteriors distribution for selected model parameters are plotted
in \autoref{fig: epsilon dot posterior} and the entire set is summarised in
\autoref{tab: epsilon dot prior posterior} along with their prior
distributions.  Comparing this to the base-model, two features are notable: the
posterior mean of $\epsilonP$ is fractionally smaller, and $\epsilondotP$ has a
posterior mean quite different from its prior, with a positive mean and
essentially zero probability of $\epsilondotP=0$. Since $\epsilondotP > 0$, the
deformation is growing with time as expected from the observation that $\Pfp$
is decreasing.
\begin{table}
\centering
\caption{\completepriorposteriorcaption{\modelEpsDot}}
\label{tab: epsilon dot prior posterior}
\scriptsize
\tabcolsep=0.07cm
\begin{tabular}{llll} \hhline{====}
        & Prior & Posterior median $\pm$ s.d. &  Units\\ \hline
$\nu_0$ & $\mathcal{N}$(2.46887171470, ${7.0}{\times} 10^{\textrm{-}11}$) & 2.47$\pm$${7.0}{\times} 10^{\textrm{-}11}$ & Hz \\
$\dot{\nu}_0$ & $\mathcal{N}$(${\text{\textrm{-}}3.658728}{\times} 10^{\textrm{-}13}$, ${5.0}{\times} 10^{\textrm{-}19}$) & ${\text{\textrm{-}}3.66}{\times} 10^{\textrm{-}13}$$\pm$${5.0}{\times} 10^{\textrm{-}19}$ & Hz/s \\
$\ddot{\nu}_0$ & $\mathcal{N}$(${8.72}{\times} 10^{\textrm{-}25}$, ${9.0}{\times} 10^{\textrm{-}27}$) & ${8.75}{\times} 10^{\textrm{-}25}$$\pm$${8.9}{\times} 10^{\textrm{-}27}$ & Hz/s$^{2}$ \\
$\epsilon_\mathrm{p}$
 & $|\mathcal{N}$(0, ${1}{\times} 10^{\textrm{-}8}$)$|$ & ${9.05}{\times} 10^{\textrm{-}9}$$\pm$${2.7}{\times} 10^{\textrm{-}11}$ &   \\
$\dot{\epsilon}_\mathrm{p}$ & $|\mathcal{N}$(0, ${2}{\times} 10^{\textrm{-}18}$)$|$ & ${1.34}{\times} 10^{\textrm{-}18}$$\pm$${5.6}{\times} 10^{\textrm{-}20}$ & s$^{\textrm{-}1}$ \\
$\theta$
 & Unif(0, 0.1) & 0.0560$\pm$0.0011 & rad \\
$\chi$
 & Unif($2\pi/5$, $\pi/2$) & 1.5529$\pm$0.0007 & rad \\
$\psi_0$
 & Unif(0, $2\pi$) & 4.7725$\pm$0.0404 & rad \\
$\rho_{2}^{0}$
 & Unif(0, 0.1464) & 0.0236$\pm$0.0003 & rad \\
$\rho_{2}''$
 & $\mathcal{N}$(0, 6.83) & 3.26$\pm$0.2 & rad$^{\textrm{-}2}$ \\
$\cos(\iota)$
 & Unif(\textrm{-}1, 1) & ${6.69}{\times} 10^{\textrm{-}3}$$\pm$${1.3}{\times} 10^{\textrm{-}3}$ &   \\
$\sigma_{\dot{\nu}}$
 & Unif(0, ${1}{\times} 10^{\textrm{-}15}$) & ${2.57}{\times} 10^{\textrm{-}16}$$\pm$${1.2}{\times} 10^{\textrm{-}17}$ & $\mathrm{s}^{\textrm{-}2}$ \\
$\sigma_{W_{10}}$
 & Unif(0, ${5.0}{\times} 10^{\textrm{-}3}$) & ${1.47}{\times} 10^{\textrm{-}3}$$\pm$${4.0}{\times} 10^{\textrm{-}5}$ & s \\
\hhline{====}
\end{tabular}

\end{table}
As pointed out earlier, we recall that due to the degeneracy of the spin-down
rate and beam-width with respect to the sign of $\epsilonP$, this should
therefore generally be interpreted as $|\epsilondotP|>0$.

\begin{figure}
\centering
\includegraphics[]{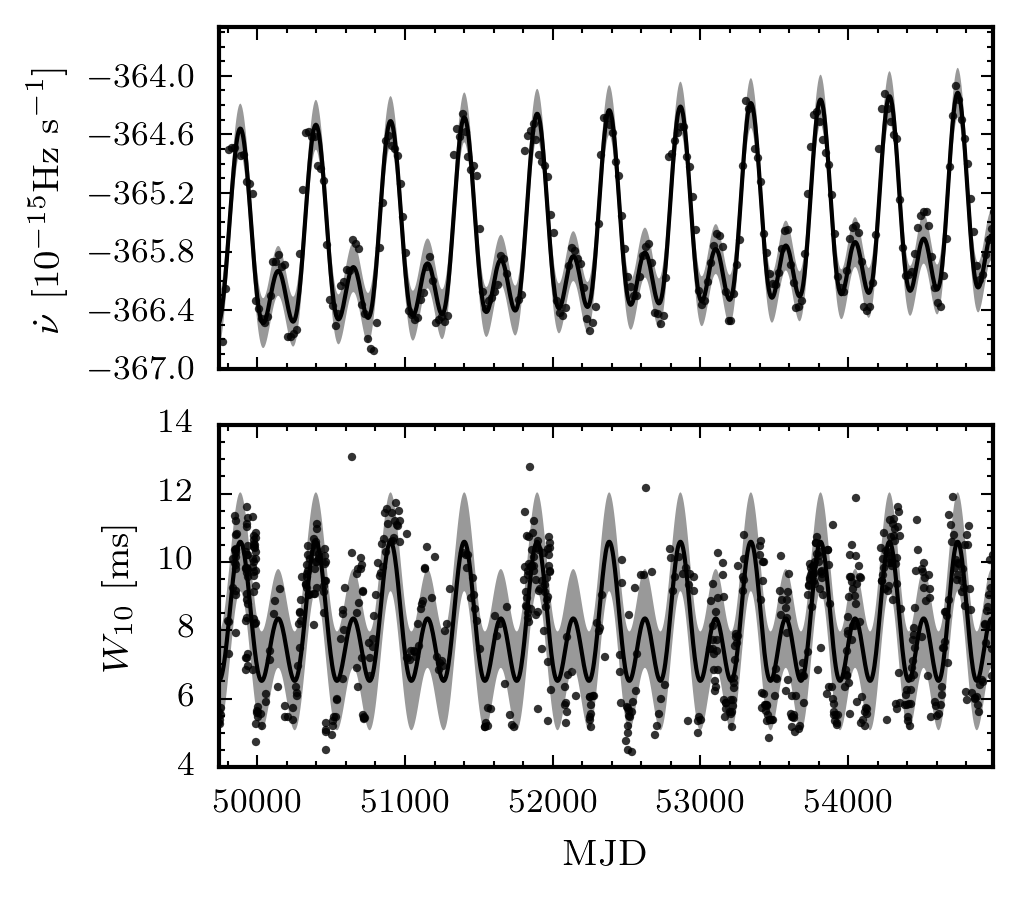}
\caption{Comparison between the MPE \modelEpsDot-model (solid line) and the
observed spin-down and beam-width data (black dots).  The shaded region
indicates the estimated $1\,\sigma$ noise level.}
\label{fig: epsilon dot posterior fit}
\end{figure}

In \autoref{fig: epsilon dot posterior fit} we plot the MPE spin-down and
beam-width functions given by the model together with the observed data.
Comparing this to \autoref{fig: base-model posterior fit} it is evident that
the model extension of \autoref{eqn: epsilon extension}, allowing for
evolution of the precession period via $\epsilondotP$, noticeably improves the
description of the data compared to the base-model.
This improvement is confirmed by the odds between the \modelEpsDot-model and
the base-model which are found as $10^{\oddsEpsilonDotBaseModel \pm
\errEpsilonDotBaseModel}$, i.e.\ decisive evidence in favour of an increasing
deformation $|\epsilonP|$ as opposed to a constant.

To understand how the two data sources contribute to the total odds, we
repeat the analysis on the two data sets independently and find that the odds
for the spin-down data are
$10^{\oddsEpsilonDotSpindownOnlyBaseModelSpindownOnly \pm
\errEpsilonDotSpindownOnlyBaseModelSpindownOnly}$ while the odds for the
beam-width data are $10^{\oddsEpsilonDotBeamwidthOnlyBaseModelBeamwidthOnly \pm
\errEpsilonDotBeamwidthOnlyBaseModelBeamwidthOnly}$ such that the individual
odds approximately sum to the combined odds. One would expect the odds to sum
up this way if the posteriors (when conditioned on each data set individually)
are consistent; we show this is the case in Appendix~\ref{app: consistency}.
The independent odds show that each data set separately strongly favours the
\modelEpsDot-model, with the (clearly much cleaner) spin-down data providing
stronger evidence than the beam-width data.

The large numerical values of the odds we obtain are related to the fact
that for a Gaussian-noise model the log-odds scale linearly with the number of
data points.  For the spin-down data set, which consisted of $257$ data points,
the average log-odds contributed by each point is $49.61/257 \approx 0.19$, or
a factor of $10^{0.19} \approx 1.6$ per data point to the odds itself.  For the
beam-width data, the corresponding numbers are $23.42/756 \approx 0.03$, or a
factor of $10^{0.03} \approx 1.07$ increase in odds per data point.  This
illustrates that it is the combination of many data points, each of which (on
average) only modestly favours the  \modelEpsDot-model, that leads to the large
overall odds.

The timescale of the inferred increase in deformation is seen to be quite
short: from the MCMC samples we calculate the median and standard deviation of
the corresponding timescale to be
\begin{align}
\tauEpsilon \equiv \frac{\epsilonP}{\epsilondotP} =
\EpsilonDottauEpsilonYearsMedian \pm
\EpsilonDottauEpsilonYearsStd\,\years.
\label{eqn: epsilon dot tau epsilon}
\end{align}

\section{Secular evolution of wobble angle and
deformation: the \modelThetaEpsDot-model}
\label{sec: evolution of the wobble angle and deformation}

In \autoref{sec: evolution of the wobble angle} we showed that variations of
$\theta$ cannot be responsible for the observed changing modulation period
$\Pfp$.  In the precession model considered here, the only plausible
explanation for the decreasing $\Pfp$ comes from allowing for an increasing
deformation $|\epsilonP|$.  However, physically it is still quite plausible for
the wobble angle $\theta$ to change over time, and at the minimum this allows
us to set limits on the rate of change of $\theta$, which has potentially
interesting implications for the crust-core coupling.  In this section, we will
therefore consider a combined extension allowing for both $\theta$ and
$\epsilonP$ to undergo linear secular evolution.  This will allow us to set
more stringent and realistic limits on the allowed $\dot{\theta}$ rates than
those provided in \autoref{sec: evolution of the wobble angle}.

\subsection{Applying the \modelThetaEpsDot-model to the data}

In order to extend the base-model with both \autoref{eqn: theta extension}
and \autoref{eqn: epsilon extension}, we simply use the same formulations
and priors as those given in \autoref{sec: evolution of the wobble angle} and
\autoref{sec: continuous evolution of epsilon}.

\begin{figure}
\centering
\includegraphics[width=0.5\textwidth]{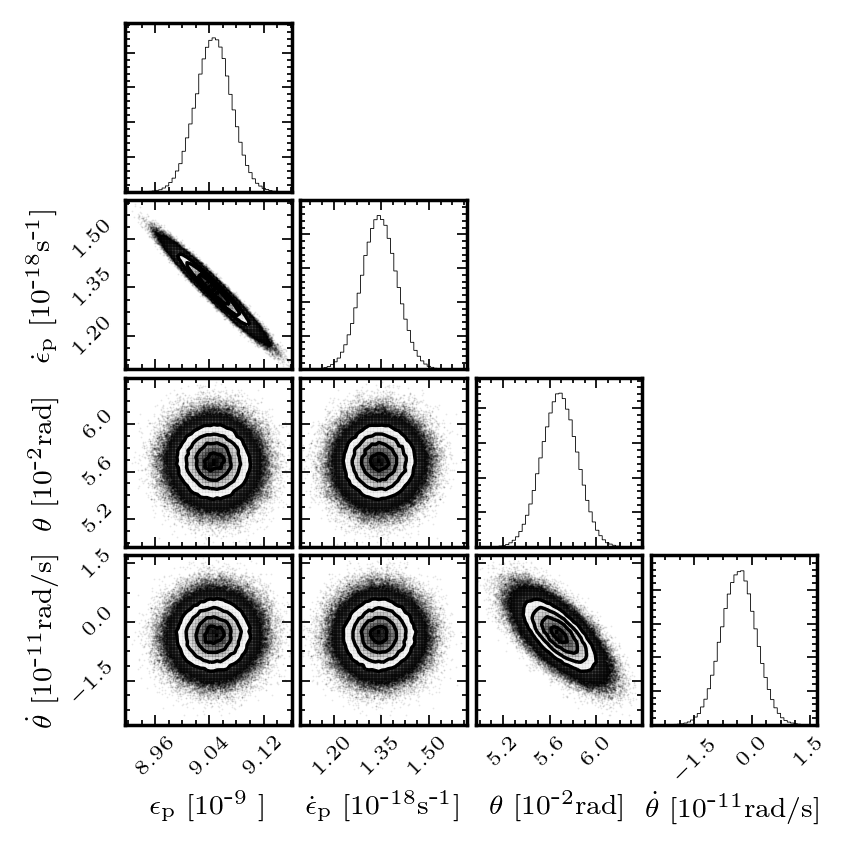}
\caption{\selectedposteriorscaption{wobble angle $\theta$, the deformation
$\epsilonP$ and their rates of change}{\modelThetaEpsDot}}
\label{fig: epsilon dot theta dot posterior}
\end{figure}

\autoref{fig: epsilon dot theta dot posterior} shows the posteriors obtained
for the deformation $\epsilonP$, the wobble angle $\theta$, and their
time-derivatives, and \autoref{tab: epsilon dot theta dot prior posterior}
summarises the posteriors found for all the model parameters.  We note that the
posterior for $\dot{\theta}$ has again a slightly negative mean, but a narrower
width than in the \modelThetaDot-model shown in \autoref{fig: theta dot
posterior}.  While the evolution in $\theta$ and $\epsilonP$ cannot be strictly
separated, the evolution of the deformation $\epsilonP$ accounts mostly for the
time varying modulation period, while the evolution of the wobble angle
$\theta$ primarily probes the variation in amplitude.
\begin{table}
\centering
\caption{\completepriorposteriorcaption{\modelThetaEpsDot}}
\label{tab: epsilon dot theta dot prior posterior}
\scriptsize
\tabcolsep=0.07cm
\begin{tabular}{llll} \hhline{====}
        & Prior & Posterior median $\pm$ s.d. &  Units\\ \hline
$\nu_0$ & $\mathcal{N}$(2.46887171470, ${7.0}{\times} 10^{\textrm{-}11}$) & 2.47$\pm$${7.0}{\times} 10^{\textrm{-}11}$ & Hz \\
$\dot{\nu}_0$ & $\mathcal{N}$(${\text{\textrm{-}}3.658728}{\times} 10^{\textrm{-}13}$, ${5.0}{\times} 10^{\textrm{-}19}$) & ${\text{\textrm{-}}3.66}{\times} 10^{\textrm{-}13}$$\pm$${4.9}{\times} 10^{\textrm{-}19}$ & Hz/s \\
$\ddot{\nu}_0$ & $\mathcal{N}$(${8.72}{\times} 10^{\textrm{-}25}$, ${9.0}{\times} 10^{\textrm{-}27}$) & ${8.75}{\times} 10^{\textrm{-}25}$$\pm$${8.9}{\times} 10^{\textrm{-}27}$ & Hz/s$^{2}$ \\
$\epsilon_\mathrm{p}$
 & $|\mathcal{N}$(0, ${1}{\times} 10^{\textrm{-}8}$)$|$ & ${9.05}{\times} 10^{\textrm{-}9}$$\pm$${2.7}{\times} 10^{\textrm{-}11}$ &   \\
$\dot{\epsilon}_\mathrm{p}$ & $|\mathcal{N}$(0, ${2}{\times} 10^{\textrm{-}18}$)$|$ & ${1.34}{\times} 10^{\textrm{-}18}$$\pm$${5.6}{\times} 10^{\textrm{-}20}$ & s$^{\textrm{-}1}$ \\
$\theta$
 & Unif(0, 0.1) & 0.0568$\pm$0.0015 & rad \\
$\dot{\theta}$ & $\mathcal{N}$(0, ${2.2}{\times} 10^{\textrm{-}10}$) & ${\text{\textrm{-}}3.38}{\times} 10^{\textrm{-}12}$$\pm$${4.5}{\times} 10^{\textrm{-}12}$ & rad/s \\
$\chi$
 & Unif($2\pi/5$, $\pi/2$) & 1.5529$\pm$0.0007 & rad \\
$\psi_0$
 & Unif(0, $2\pi$) & 4.7703$\pm$0.0398 & rad \\
$\rho_{2}^{0}$
 & Unif(0, 0.1464) & 0.0236$\pm$0.0004 & rad \\
$\rho_{2}''$
 & $\mathcal{N}$(0, 6.83) & 3.25$\pm$0.2 & rad$^{\textrm{-}2}$ \\
$\cos(\iota)$
 & Unif(\textrm{-}1, 1) & ${6.77}{\times} 10^{\textrm{-}3}$$\pm$${1.3}{\times} 10^{\textrm{-}3}$ &   \\
$\sigma_{\dot{\nu}}$
 & Unif(0, ${1}{\times} 10^{\textrm{-}15}$) & ${2.57}{\times} 10^{\textrm{-}16}$$\pm$${1.2}{\times} 10^{\textrm{-}17}$ & $\mathrm{s}^{\textrm{-}2}$ \\
$\sigma_{W_{10}}$
 & Unif(0, ${5.0}{\times} 10^{\textrm{-}3}$) & ${1.47}{\times} 10^{\textrm{-}3}$$\pm$${4.0}{\times} 10^{\textrm{-}5}$ & s \\
\hhline{====}
\end{tabular}

\end{table}

\begin{figure}
\centering
\includegraphics[width=0.5\textwidth]{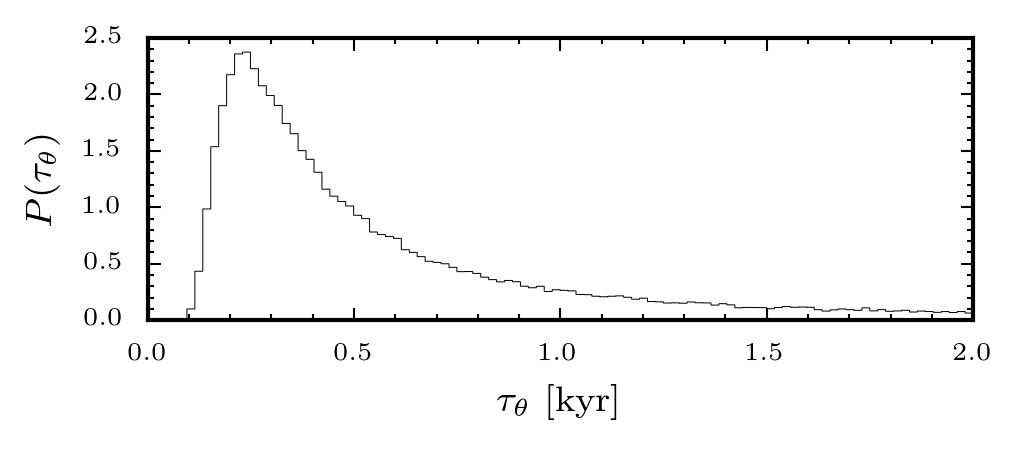}
\caption{The posterior distribution for the $\theta$-evolution timescale
$\tauTheta=|\theta/\dot{\theta}|$ under the \modelThetaEpsDot-model.}
\label{fig: EpsilonDotThetaDot tau Theta}
\end{figure}

\autoref{fig: EpsilonDotThetaDot tau Theta} shows the resulting posterior
for the timescale of $\theta$-evolution, $\tauTheta = |\theta / \dot{\theta}|$.
We see that the tighter posterior on $\dot{\theta}$ shifts the probability of
$\tauTheta$ to larger values than those seen in \autoref{fig: theta dot tau
Theta}, favouring slower rates of change of $\theta$.

We can place a 95\% credible lower limit of $\tauTheta >
\EpsilonDotThetaDottauThetaFiveYears\,\years$ and the distribution has a median
value of $\EpsilonDotThetaDotPtauThetaMedianYears\,\years$.  In this combined
model, $\tauEpsilon=\EpsilonDotThetaDottauEpsilonMedianYears \pm
\EpsilonDotThetaDottauEpsilonStdYears\,\years$ (the timescale remains unchanged
from the \modelEpsDot-model considered in \autoref{sec: continuous evolution
of epsilon}).

We obtain the odds in favour of the \modelThetaEpsDot-model compared to the
base-model as $10^{\oddsEpsilonDotThetaDotBaseModel \pm
\errEpsilonDotThetaDotBaseModel}$, i.e.\ slightly less than for the
\modelEpsDot-model.  We see that, similarly to the case of the
\modelThetaDot-model, the introduction of $\dot{\theta}$ does not produce a
significant-enough improvement in the fit compared to the increase in prior
volume.

\section{Discrete jumps in deformation: the \modelEpsJumps-model}
\label{sec: resolved discrete jumps in epsilon}

The success of the \modelEpsDot-model of \autoref{sec: continuous evolution of
epsilon} indicates that a time-dependent $\epsilonP(t)$ provides a significant
improvement over the base-model.  In this section, we explore an alternative to
the slow secular change by modelling the time-variation as a set of discrete
jumps in $\epsilonP$.

\subsection{Defining the \modelEpsJumps-model}
In this model extension, we allow $\epsilonP$ to undergo $N$ distinct
\emph{positive} jumps.  For each jump $j \in[1,\,N]$ at time $t_j$, we define
two dimensionless parameters: the fractional observation time at which the jump
occurs, $R_{j} \equiv (t_j - t_0)/\Tobs\in [0, 1]$, where $t_0$ is the
start-time and $\Tobs$ is the total observation time, and the fractional
(positive) variation in $\epsilonP$ at the jump, $\Delta_{j} \equiv
\Delta\epsilon_{\mathrm{p},j}/\epsilon_{\mathrm{p},0}\in [0,\infty)$.  In this
way, the time evolution of $\epsilonP(t)$ can be written as
\begin{equation}
\epsilonP(t) = \epsilon_{\mathrm{p},0} \, \left(1
+ \sum_{j=1}^{N}H(t - t_{0} - R_{j}\,\Tobs)\,\Delta_{j}\right),
\label{eqn: epsilonP delta}
\end{equation}
where $H(t)$ is the Heaviside step function.

\subsection{Applying the \modelEpsJumps-model to the data}

We assign a uniform prior distribution over the total observation span for
$R_{j}$, the time of the jumps, with $R_{j} < R_{j+1} \; \forall \; j$. For the
jump sizes $\Delta_{j}$ we will use a prior consistent with the
\modelEpsDot-model (see \autoref{sec: continuous evolution of epsilon}),
specifically a zero-mean Gaussian for $\epsilondotP$ with standard-deviation of
$2\times10^{-18}$~s$^{-1}$. Distributing an equivalent total change in
$\epsilonP$ on $N$ discrete jumps, this gives an approximate scale of
\begin{align}
\Delta \approx \frac{\epsilondotP \, \Tobs}{\epsilonP\,N} \approx \frac{0.1}{N},
\end{align}
where we have substituted $\epsilonP$ and
$\epsilondotP$ for the prior standard-deviation used in the \modelEpsDot-model.
We use this to set the scale for a Gaussian prior on the fractional jump size
as $\Delta_j \sim |\mathcal{N}(0,\,0.1/N)|$.

To speed up the fitting process we have modified the original MCMC fitting
process described in Appendix~A of \citet{ashton2016}. Specifically, it was
found that when fitting for the jump parameters, the MCMC chains took a long
time to find the base-model best estimates for the spin-down parameters $\nuo$,
$\nudo$, and $\nuddo$ and the angles $\chi$ and $\theta$. Therefore, instead of
initialising the chains from the prior, for the parameters shared with the
base-model we initialise them from the base-model posterior. This modification
does not change our final estimates, provided that the burn-in period is
sufficiently long to allow them to evolve from this point and explore all areas
of the parameter space. For several values of $N$, we tested that evolving from
the prior produced the same results, but the computation took longer to
converge.

The number of jumps $N$ can itself be thought of as a model parameter: ideally
we would fit $N$ as part of the MCMC sampling. However, to do this one must use
a reversible-jump MCMC algorithm which can vary the number of model dimensions.
This is not currently implemented in the software used in this analysis.
Instead we have opted for a crude, but sufficient method in which we fit the
model for different values of $N$ individually and then use the respective odds
to compare them.  For each increase in $N$, the number of steps required to
reach convergence increases.
\begin{figure}
\centering
\includegraphics[width=.4\textwidth]{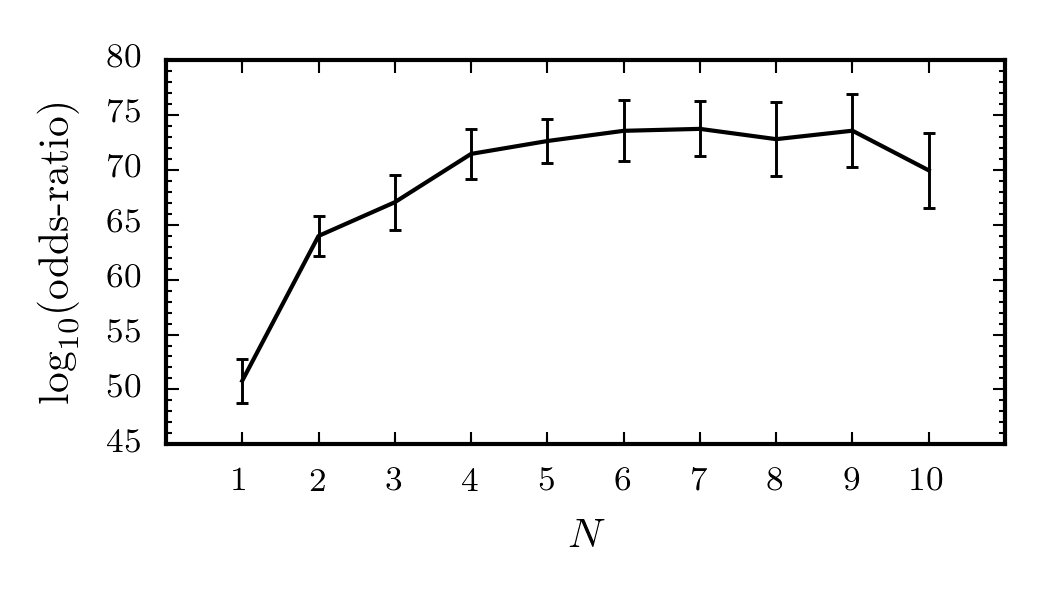}
\caption{The log odds-ratio for the \modelEpsJumps-model
for a varying number of jumps $N$ compared to the base-model.}
\label{fig: log odds N}
\end{figure}
In \autoref{fig: log odds N} we show the odds of the $N$-jump model compared
to the base-model as a function of the number of jumps $N$. We see that up to
$N\sim6$ the odds increase, then reach a plateau and start to marginally
decrease for $N=10$.
\begin{figure}
\centering
\includegraphics[]{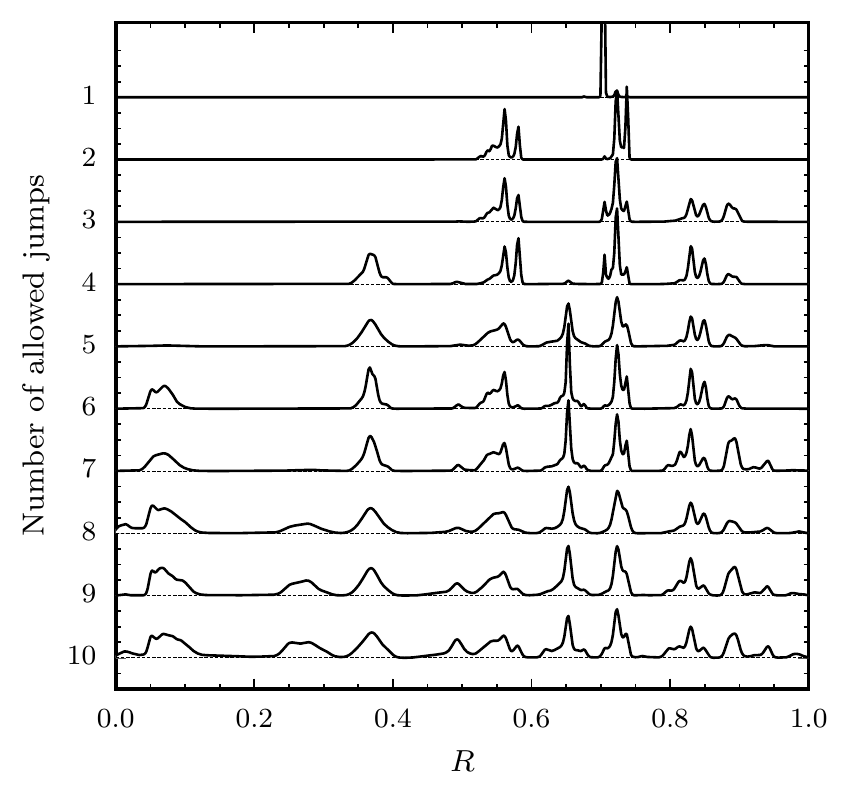}
\caption{The combined posteriors for the fractional jump times $R$ for the
different $N$-jump \modelEpsJumps-models.  The left-hand axis indicates the
respective number of jumps $N$. For each $N$, a vertical offset has been added
to each posterior to allow them to be distinguished and dashed lines mark
the `zero' line.}
\label{fig: RStacked}
\end{figure}
In \autoref{fig: RStacked} we present a stacked plot showing the posteriors on
the jump times $R$ for all jumps, for the different $N$-jump models. For ease
of reading the plot, each jump is normalised so that the area under the $N=1$
line is 1, under the $N=2$ model the area is 2, etc.

The positions $R$ at which the jumps occur appear consistent \emph{between}
different $N$-jump models.  Moreover, the posteriors for each jump are
multimodal, each having a unique `fingerprint', which also appear consistent
between models.  This would not necessarily be expected if the best fit was
quite agnostic about the exact jump times and simply distributed $N$ jumps
randomly over the observation period.  We also see a consistent progression
play out as the number of allowed steps $N$ is increased: up to $N=6$ each
increase in $N$ finds a new jump site, but from $N\ge7$ the new jump sites are
not so well defined. However, we cannot rule out the
possibility that the MCMC chains did not successfully converged for some of
these models.

The data does not seem to strongly favour a particular number of jumps above
$N\ge6$. Therefore, for illustrative purposes we will use $N=6$ as our
posterior estimate for $N$. While this model does not have the largest
odds-ratio (as shown in \autoref{fig: log odds N}), the difference to the N=7
model, which does have the largest odds-ratio, is much smaller than the error
bars. Moreover, this model captures all of the essential features of the
discrete jumps as seen in \autoref{fig: RStacked}.

\subsection{The $N=6$ \modelEpsJumps-model}

\begin{figure}
\centering
\includegraphics[width=0.5\textwidth]{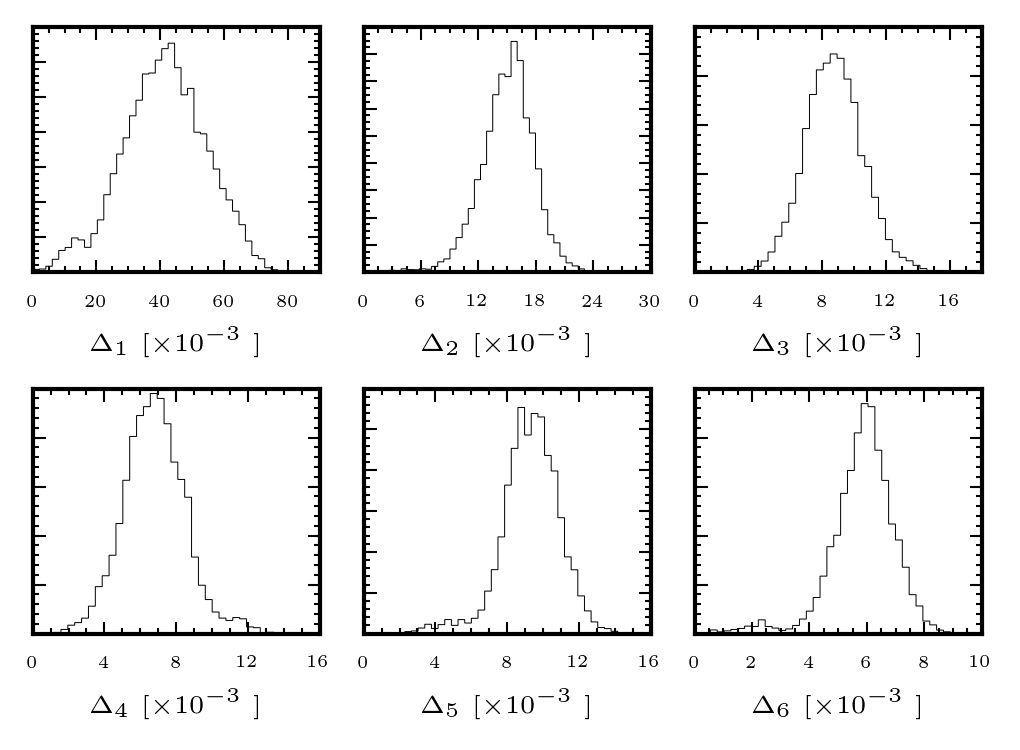}
\caption{\selectedposteriorscaption{six relative jump-size parameters
         $\Delta_j$}{\modelEpsJumps}}
\label{fig: epsilon delta posterior}
\end{figure}

\begin{figure}
\centering
\includegraphics[]{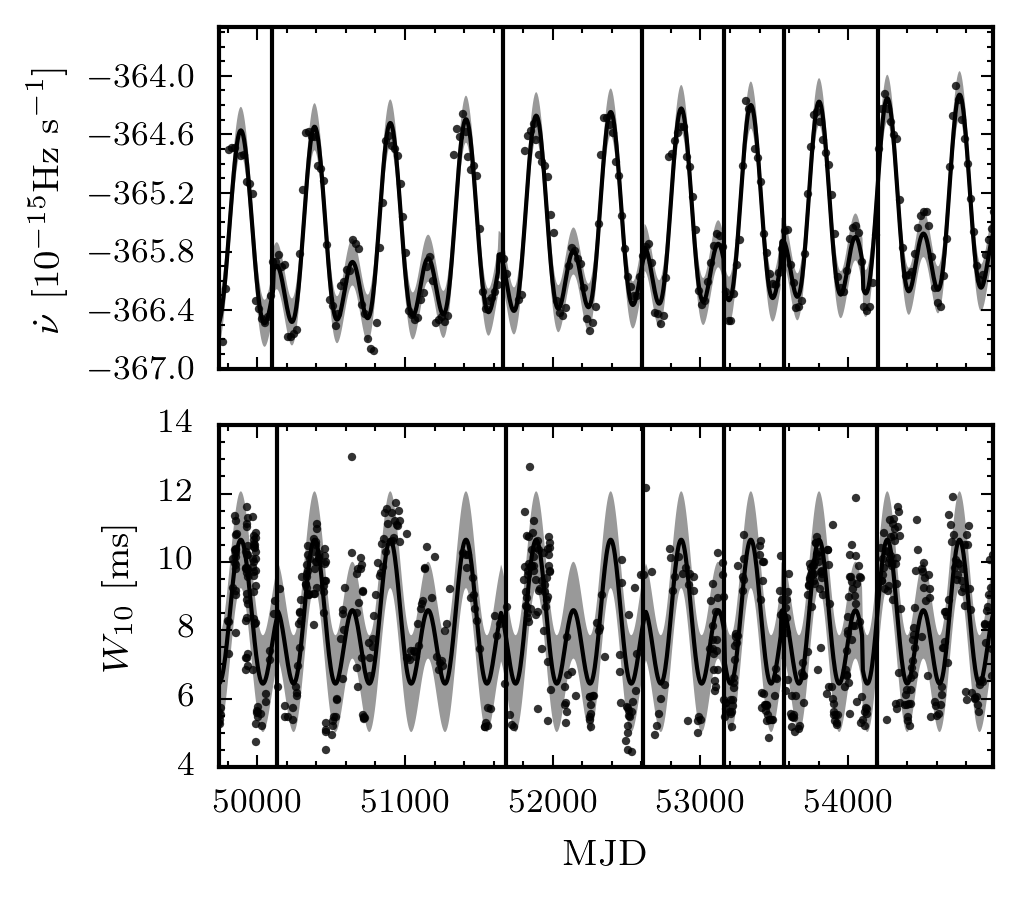}
\caption{Comparison between the MPE \modelEpsJumps-model with $N=6$ jumps
(solid line) and the observed spin-down and beam-width data (black dots).
Vertical lines mark the times of the six (positive) jumps in $\epsilonP(t)$.
The shaded region indicates the estimated $1\,\sigma$ noise level.}
\label{fig: epsilon delta posterior fit}
\end{figure}

\autoref{fig: epsilon delta posterior} shows the posterior for the six
relative jump sizes $\Delta_{j}$ which have typical sizes of order
$\Delta_j\sim 0.01$. We provide a summary of the priors and posteriors for all
the model parameters in \autoref{tab: epsilon delta prior posterior}.  Then,
in \autoref{fig: epsilon delta posterior fit} we show the MPE fits to the
spin-down and beam-width data; we indicate the jump times with vertical lines.
By eye, the fit shows a similar level of improvement compared to the base-model
\autoref{fig: base-model posterior fit} as that observed in \autoref{fig:
epsilon dot posterior fit}, which is consistent with the similar odds of
$10^{\oddsEpsilonDeltasixJumpsBaseModel \pm \errEpsilonDeltasixJumpsBaseModel}$
relative to the base-model. As such we cannot distinguish between the two types
of evolving deformation (continuous evolution verses discrete jumps).

\begin{table}
\centering
\caption{\completepriorposteriorcaption{$N=6$ \modelEpsJumps}}
\label{tab: epsilon delta prior posterior}
\scriptsize
\tabcolsep=0.07cm
\begin{tabular}{llll} \hhline{====}
        & Prior & Posterior median $\pm$ s.d. &  Units\\ \hline
$\nu_0$ & $\mathcal{N}$(2.46887171470, ${7.0}{\times} 10^{\textrm{-}11}$) & 2.47$\pm$${7.2}{\times} 10^{\textrm{-}11}$ & Hz \\
$\dot{\nu}_0$ & $\mathcal{N}$(${\text{\textrm{-}}3.658728}{\times} 10^{\textrm{-}13}$, ${5.0}{\times} 10^{\textrm{-}19}$) & ${\text{\textrm{-}}3.66}{\times} 10^{\textrm{-}13}$$\pm$${4.7}{\times} 10^{\textrm{-}19}$ & Hz/s \\
$\ddot{\nu}_0$ & $\mathcal{N}$(${8.72}{\times} 10^{\textrm{-}25}$, ${9.0}{\times} 10^{\textrm{-}27}$) & ${8.75}{\times} 10^{\textrm{-}25}$$\pm$${8.6}{\times} 10^{\textrm{-}27}$ & Hz/s$^{2}$ \\
$\epsilon_\mathrm{p}$
 & $|\mathcal{N}$(0, ${1}{\times} 10^{\textrm{-}8}$)$|$ & ${8.82}{\times} 10^{\textrm{-}9}$$\pm$${1.4}{\times} 10^{\textrm{-}10}$ &   \\
$\Delta_{1}$ & $|\mathcal{N}$(0, 0)$|$ & 0.04$\pm$0.0 &  \\
$R_{1}$ & Unif(0, 1) & 0.07$\pm$0.0 &  \\
$\Delta_{2}$ & $|\mathcal{N}$(0, 0)$|$ & 0.02$\pm$${2.8}{\times} 10^{\textrm{-}3}$ &  \\
$R_{2}$ & Unif(0, 1) & 0.37$\pm$0.0 &  \\
$\Delta_{3}$ & $|\mathcal{N}$(0, 0)$|$ & ${8.71}{\times} 10^{\textrm{-}3}$$\pm$${1.8}{\times} 10^{\textrm{-}3}$ &  \\
$R_{3}$ & Unif(0, 1) & 0.55$\pm$0.0 &  \\
$\Delta_{4}$ & $|\mathcal{N}$(0, 0)$|$ & ${6.76}{\times} 10^{\textrm{-}3}$$\pm$${1.7}{\times} 10^{\textrm{-}3}$ &  \\
$R_{4}$ & Unif(0, 1) & 0.65$\pm$0.0 &  \\
$\Delta_{5}$ & $|\mathcal{N}$(0, 0)$|$ & ${9.37}{\times} 10^{\textrm{-}3}$$\pm$${1.7}{\times} 10^{\textrm{-}3}$ &  \\
$R_{5}$ & Unif(0, 1) & 0.73$\pm$0.0 &  \\
$\Delta_{6}$ & $|\mathcal{N}$(0, 0)$|$ & ${5.94}{\times} 10^{\textrm{-}3}$$\pm$${1.1}{\times} 10^{\textrm{-}3}$ &  \\
$R_{6}$ & Unif(0, 1) & 0.85$\pm$0.0 &  \\
$\theta$
 & Unif(0, 0.1) & 0.0572$\pm$0.0011 & rad \\
$\chi$
 & Unif($2\pi/5$, $\pi/2$) & 1.5539$\pm$0.0007 & rad \\
$\psi_0$
 & Unif(0, $2\pi$) & 4.7351$\pm$0.0535 & rad \\
$\rho_{2}^{0}$
 & Unif(0, 0.1464) & 0.0233$\pm$0.0003 & rad \\
$\rho_{2}''$
 & $\mathcal{N}$(0, 6.83) & 3.33$\pm$0.2 & rad$^{\textrm{-}2}$ \\
$\cos(\iota)$
 & Unif(\textrm{-}1, 1) & ${7.2}{\times} 10^{\textrm{-}3}$$\pm$${1.2}{\times} 10^{\textrm{-}3}$ &   \\
$\sigma_{\dot{\nu}}$
 & Unif(0, ${1}{\times} 10^{\textrm{-}15}$) & ${2.44}{\times} 10^{\textrm{-}16}$$\pm$${1.2}{\times} 10^{\textrm{-}17}$ & $\mathrm{s}^{\textrm{-}2}$ \\
$\sigma_{W_{10}}$
 & Unif(0, ${5.0}{\times} 10^{\textrm{-}3}$) & ${1.44}{\times} 10^{\textrm{-}3}$$\pm$${3.9}{\times} 10^{\textrm{-}5}$ & s \\
\hhline{====}
\end{tabular}

\end{table}

\section{Interpreting the upper limit on $\dot\theta$}
\label{sec: theta_dot_interpretation}

Dissipative processes internal to the star may damp the wobble motion, leading
to a decrease in $\theta$. Looking at the posterior on $\dot\theta$ shown in
\autoref{fig: epsilon dot theta dot posterior}, we see that, while the peak
of the probability distribution lies at a value $\dot\theta < 0$,  the peak is
nevertheless close to $\dot\theta = 0$, so there is no clear evidence for any
evolution in the wobble angle over the duration of these observations.
Slightly more informatively, in \autoref{fig: EpsilonDotThetaDot tau Theta} we
plotted the posterior on the timescale $\tau_\theta = |\theta / \dot\theta|$.

Even though this analysis finds no evidence for a secular variation in the
wobble angle, we can use these results to put a lower bound on the timescale
on which $\tauTheta$ evolves, i.e.\ we can place a 95\% credible interval that
$\tauTheta > \EpsilonDotThetaDottauThetaFiveYears$~years.

Mutual friction, a dissipative coupling of neutron vortices and the
charged component of the star, is the leading candidate for damping precession.
The effect of mutual friction on precession was examined by
\citet{sedrakian1999} and \citet{glampedakis2008stability,
glampedakis2009superfluid}.  The strength of the interaction can be
parameterised by a dimensionless quantity $\mathcal{R}$, a measure of the
relative strength of the mutual friction force to the Magnus force.  In the
limit of large $\mathcal{R}$, the vortices become pinned to the crust, and a
very fast precession frequency is obtained, in contradiction with the
observations.   The free precession observation instead requires the weak drag
limit, $\mathcal{R} \ll 1$, to apply.  The damping time can be shown to be
given by
\begin{align}
\tau_{\rm MF} = \frac{1}{{\cal R} \epsilonP 2\pi \nu} \frac{I_{\rm prec}}{I_{\rm SF}} ,
\end{align}
where $I_{\rm SF}$ denotes the moments of inertia of the core superfluid (see
\citet{sedrakian1999} and Appendix A of \citet{glampedakis2009superfluid}).
Strictly, $\cal R$ is a locally-defined quantity, i.e.\ a function of density,
but this dependence is `averaged-out' in the rigid-body dynamics analysis
through which the above equation is obtained.

Given that the value of $\epsilonP$ is known from our posterior estimate, we
can, as described in  \citet{glampedakis2009superfluid}, convert
our lower bound on $\tau_\theta$ to a 95\% credible upper bound on $\cal R$
assuming that $\tauTheta = \tau_{\rm MF}$:
\begin{align}
{\cal R} \lesssim 1.2 {\times} 10^{-4}
         \left(\frac{\EpsilonDotThetaDottauThetaFiveYears \, \rm yr}{\tau_{\theta}}\right)
         \left(\frac{9.7{\times}10^{-9}}{\epsilonP}\right)
         \left(\frac{I_{\rm prec}/I_{\rm SF}}{0.1}\right) .
\end{align}
Again as noted in \citet{glampedakis2009superfluid}, this can be combined with
a lower bound on $\cal R$ that comes from analysis of the Christmas 1988 glitch
in the Vela pulsar, where the relevant coupling time can be shown to be given
by $\tau_{\rm MF} = 1/(4\pi \nu {\cal R}) I_{\rm prec}/I_{\rm SF}$.  From the
analysis of the Vela glitch by \citet{abney1996observational}, if we set
$I_{\rm prec}/I_{\rm SF} = 0.1$, we obtain $30$ seconds as the upper limit on
the crust-core coupling timescale, leading to a lower bound  ${\cal R} \gtrsim
2.4 \times 10^{-5}$. Combining these results, we have
\begin{align}
2.4 \times 10^{-5} \lesssim   {\cal R} \lesssim 1.2 \times 10^{-4} .
\end{align}
The upper limit given here is an improvement by about one order of magnitude on
that given by \citet{glampedakis2009superfluid}.

A number of authors have attempted first-principles microphysical calculations
of this parameter, appropriate for a neutron superfluid core \citep{ alpar1984,
alpar1988, andersson2006}. Taking Equation~(64) of \citet{andersson2006},
and setting the density $10^{14}$ g cm$^{-3}$, and the proton density
fraction to $0.1$, one obtains a range for ${\cal R} \approx 9.7 \times 10^{-5}$--$3.18 \times
10^{-4}$ as one varies the proton effective mass over the interval $0.5$--$0.7$
times the bare mass.  Clearly, there is a reasonable level of convergence
between the shrinking observation range in $\cal R$ reported above and
microscopic estimates.

\section{Interpreting the evolving deformation}
\label{sec: interpretation}

The rather rapid observed decrease in the free precession period is not easy to
explain within the precessional model.  We have shown above that it corresponds
to an increase in the deformation parameter $\epsilonP$ of
\autoref{eq:epsilon_P_definition}.  Re-writing this slightly,
\begin{equation}
\epsilonP = \frac{\Delta I_{\rm d}}{I_{\rm prec}} = \frac{\Delta I_{\rm d}}{I_*}   \frac{I_*}{I_{\rm prec}} ,
\end{equation}
we see that we can interpret our observation as an increase on the deformation
$\Delta I_{\rm d}/ I_*$, and/or a decrease in the fraction of the star that
participates in the free precession, $I_{\rm prec} / I_*$.  The total variation
must correspond to a timescale of $\approx 213$ years, a rather short timescale
for a $\sim 10^5$ year old neutron star.

It is difficult to motivate a variation in $I_{\rm prec} / I_*$ on this sort of
timescale.  One possible  mechanism for producing a decrease in this quantity
would be if the core superfluid does not contribute to $I_{\rm prec}$.  Then, if the
star is currently cooling through the density-dependent normal matter--superfluid matter transition, the amount of core superfluid matter will be gradually increasing, with a corresponding decrease in the amount of core normal matter, hence, by our current assumption, decreasing $I_{\rm prec}$.  Such a
mechanism has been used by \citet{ho_andersson_2012} to explain the $n < 3$
braking indices in some young pulsars.  However, it is difficult to countenance
such a mechanism applying here.  PSR~B1828-11 is a relatively old pulsar, and
probably cooled through the normal fluid/superfluid transition when it was much
younger.  Also, its observed braking index is $n \approx 16$ (see Table \autoref{tab: ATNF}), so does not have
$n <3$ as would be expected if the electromagnetic spin-down torque were acting
on a progressively smaller fraction of the stellar moment of inertia.  Also, in
the model of \citet{ho_andersson_2012}, the newly created superfluid is
required to pin to the crust, something which would result in a much more rapid
rise in the free precession frequency via the gyroscopic effect of a pinned
superfluid in a rotating star \citep{shaham1977}--see the discussion below.

The alternative possibility is that the deformation $\Delta I_{\rm d} / I_*$ is
steadily increasing.  The deformation itself may be supported by elastic and/or
magnetic strains.  In the case of elastic strains, it is very difficult to
understand why the deformation should  increase with time.  Elastic strains can
be expected to be steadily reduced by plastic flow (and possibly by occasional
crustquakes), which would lead to a decreasing deformation.

In the case of magnetically-sustained deformations, it is again puzzling that
the deformation should increase with time, as magnetic fields can be expected
to decay, although the interplay of Ohmic decay, Hall drift and ambipolar
diffusion processes can lead to a complicated evolution, with the (local) field
strength increasing in some places.  Nevertheless, the required evolution
timescale $\sim 200$ years is short compared to the timescales expected for
these processes (see e.g.\ \citet{goldreich_reisenegger_1992}).

Note that if the \emph{exterior} magnetic field also
evolves on this time scale, then we should be able to measure it from the
braking index. That is we allow $B{=}B(t)$ in the usual vacuum dipole braking
law \citep{shapiro83} and solve for the derived braking index, giving
\begin{align}
n = \frac{\ddot{\nu}\nu}{\dot{\nu}^{2}} = 3 + 2 \frac{\tauAge}{\tau_{B}}
\approx 10^3
\left(\frac{\tauAge}{10^{5}\;\mathrm{ yrs}}\right)
\left(\frac{\tau_B}{200\;\mathrm{ yrs}}\right)^{-1} .
\end{align}
This is much larger than the measured value of $n\approx 16$ (see \autoref{tab: ATNF}). So we can exclude models where the
exterior field evolves in tandem with the internal one, but it remains unclear
if the internal field could vary on such a timescale.

The possibility of the star containing a pinned superfluid component adds an
additional strand to this story.  As shown by \citet{shaham1977}, a pinned
superfluid has a profound effect on the precession frequency, adding a term
proportional to $I_{\rm PSF}$, the amount of pinned superfluid:
\begin{equation}
\label{eq:e_p_PSF}
\frac{P}{P_{\rm fp}} = \frac{\Delta I_{\rm d}}{I_{\rm prec}} + \frac{I_{\rm PSF}}{I_{\rm prec}}  ,
\end{equation}
valid for small wobble angle and with the pinning directed along the symmetry axis of the biaxial star.
Assuming that the quantity $\Delta I_{\rm d} / I_{\rm prec}$ is positive (or
else negligible), this immediately translates into the bound $I_{\rm PSF} /
I_{\rm prec} \lesssim 10^{-8}$ for PSR~B1828-11, much less than the value
expected on the basis of microphysical considerations and  superfluid glitch
theory \citep{jones2001, link2001}.  A possible explanation for this has been
advanced by \citet{linkcutler2002}, who argued that the precessional motion
itself might cause most/all of the pinning to break.

This has motivated most models of PSR~B1828-11 assuming that $I_{\rm PSF}$ is
exactly zero.  However, as noted above, a small amount of pinning is allowed.
This suggests an alternative mechanism to explain the evolving precession
period:  the previously broken pinning may be gradually re-establishing itself,
with the amount of pinned superfluid increasing steadily over the last $\sim
200$ years.  Indeed, we can estimate the timescale $\Delta t_{\textrm{re-pin}}$  for
the gradual re-pinning to re-establish a reservoir of pinned superfluid of
moment of inertia $\Delta t_{\textrm{re-pin}}$.  From \autoref{eq:e_p_PSF} we
have $\dot I_{\rm PSF} = I_{\rm prec} \dot \epsilon_{\rm p}$,  so
\begin{equation}
\Delta t_{\textrm{re-pin}} = \frac{\Delta I_{\textrm{re-pin}}}{\dot I_{\rm PSF}}
=
2.13 \times 10^8 {\, \rm yr} \,\,\, \frac{\Delta I_{\textrm{re-pin}}/I_*}{10^{-2}}
     \frac{I_*}{I_{\rm prec}}    ,
\end{equation}
implying that such unpinning events have to be rare in the pulsar population, as PSR~B1828-11 will not
build up a typically sized pinned superfluid reservoir (at the few percent
level) for a long time to come.

The ideas discussed here (evolving strain and pinned superfluidity) are all
relevant to the physics of pulsar glitches.  In fact, PSR~B1828-11 was observed
to glitch in 2009: see  \citet{espinoza2011} and
\url{www.jb.man.ac.uk/~pulsar/glitches/gTable.html}).  The interplay between
the modelling of the free precession and the glitch is an interesting topic in
its own right.  We have explored the consistency requirements between the free
precession interpretation of the observed quasi-periodicities and glitches in a
separate publication \citep{letter}, which exposes significant tensions between
the small wobble angle free precession model considered here and standard
models of pulsar glitches.

\section{Discussion \& Outlook}
\label{sec: discussion}

In this work, we have extended the free precession model of \citet{ashton2016}
to allow for both the wobble angle $\theta$ and the deformation $\Delta I_{\rm
d}/I_{\rm prec}$ of PSR~B1828-11 to evolve  in time.  The generalisation to
allow for $\theta$ to vary was extremely natural, as dissipative processes
internal to the star are expected to affect the wobble angle, causing it to
decay in oblate stars ($\Delta I_{\rm d} >0$), and grow in prolate ones
($\Delta I_{\rm d} <0$; \citep{cutler2002}).  That the deformation $\Delta
I_{\rm d}/I_{\rm prec}$ should vary in time is less obvious.  However, we first
showed, in a completely model independent way (i.e.\ independently of the cause
of the quasi-periodic oscillation in spin-down rate) that the $\sim 500$ day
modulation period was getting shorter; this necessitated the allowance for a
time-varying deformation in our precession model.

We in fact found no evidence for a variation in the wobble angle, with the
inclusion of this new effect not producing a significant improvement in our
ability to fit the data.  We therefore proceeded to set an upper limit on the
timescale on which it varied, $\tau_\theta \gtrsim 171$ years.  We translated
this into an upper bound on the strength of the mutual frictional parameter
${\cal R} \lesssim 1.2 \times 10^{-4}$, describing the strength of the coupling
between the crust and core, improving on previously published results by
approximately one order of magnitude.  When combined with a lower limit on the
strength of this coupling, as deduced previously by analysis of the Vela 1988
glitch, this parameter is confined to the interval $2.4 \times 10^{-5} \lesssim
{\cal R} \lesssim 1.2 \times 10^{-4}$, a rather narrow range, but consistent
with microscopic calculation.

In terms of the evolving deformation, we explored two phenomenological ways to
model this: either as a smooth secular evolution of the deformation or as $N$
discrete jumps in the deformation.  We find that both of these models
produce a substantial improvement in the fit when compared to the
base-model---decisive evidence that, in the context of precession, the
magnitude of the star's deformation is growing; this can be seen in
\autoref{tab: log odds} where we list the odds-ratios for all models
extensions considered in this work.  For the discrete jumps model discussed in
\autoref{sec: resolved discrete jumps in epsilon}, we found $6$ or more jumps
seemed to produce the best fit and used the $N=6$ model to illustrate our results.
\begin{table}
\centering
\caption{The log odds-ratio for all tested models against the base-model.}
\label{tab: log odds}
\begin{tabular}{lc}
Model & $\log_{10}(\mathrm{odds-ratio})$\\ \hline
\modelThetaDot &
$ \oddsThetaDotBaseModel \pm \errThetaDotBaseModel $\\
\modelEpsDot &
$ \oddsEpsilonDotBaseModel \pm \errEpsilonDotBaseModel $\\
\modelThetaEpsDot &
$ \oddsEpsilonDotThetaDotBaseModel \pm \errEpsilonDotThetaDotBaseModel $\\
N=6 \modelEpsJumps &
$ \oddsEpsilonDeltasixJumpsBaseModel \pm \errEpsilonDeltasixJumpsBaseModel $ \\
\end{tabular}
\end{table}

The odds-ratio between the \modelEpsDot-model and the N=6 \modelEpsJumps-model
is $10^{\oddsEpsilonDotEpsilonDeltasixJumps\pm
\errEpsilonDotEpsilonDeltasixJumps}$, so  we find no evidence to favour one of
these two evolution models over the other. For both models an approximately
equivalent informative prior was used, but when the odds-ratio is marginal the
prior can have a substantial effect. We therefore cannot state without further
investigation which of the two model extensions is preferred with certainty and
without unfounded bias from the prior. It would be useful to propose
substantive physical models which have well defined priors; this would allow a
more thorough statement to be made.

We discussed the possible physical cause of the evolution in the deformation.
We mentioned elastic, magnetic and pinned superfluid interpretations, and
pointed out some difficulties with all of these.  PSR~B1828-11 underwent a
glitch in 2009 \citep{espinoza2011}.  In a separate publication, we discuss
consistency requirements between the precession model described here and the
glitch, folding in the evolving precession period into our discussion
\citep{letter}.

In interpreting this changing deformation, it may be important to note that
while in this analysis we fitted the `small-$\chi$' model (as defined by
\citet{arzamasskiy2015}), our analysis can equally be applied to the
`large-$\chi$' model by interchange of the $\theta$ and $\chi$ labels at the
parameter estimation stage. This is shown in Appendix~\ref{app: derivation} and
is due to the symmetry in $\theta$ and $\chi$ in the spin-down and beam-width
models. The two solutions correspond to quite different physical scenarios
which may result in fundamental differences in their interpretation.

The findings presented in this work provide a new way to probe neutron star
physics.  It remains to be understood what is the true cause of the changing
deformation and whether this happens as a smooth secular evolution or as a
number of discreet jumps.  Moreover, it would be interesting to know if
alternative models to precession can better model this behaviour.

\section*{Acknowledgements} GA acknowledges financial support from the
University of Southampton and the Albert Einstein Institute (Hannover).  DIJ
acknowledges support from STFC via grant number ST/H002359/1, and also travel
support from NewCompStar (a COST-funded Research Networking Programme).  We thank Nils Andersson for comments on this manuscript, Betina Posselt for advice on setting limits on X-ray flux,
\citet{dan_foreman_mackey_2014_11020} for the software used in generating
posterior probability distributions, and \citet{lyne2010} for generously
sharing the data for PSR~B1828-11.

\bibliographystyle{mnras}
\bibliography{bibliography}

\appendix
\section{Derivation of the spin-down rate and the $\theta \leftrightarrow \chi$
symmetry}
\label{app: derivation}
In this appendix, we derive the spin-down rate for a precessing pulsar under a
vacuum point-dipole spin-down torque. We will use a generalisation of vacuum
point-dipole torque to allow for a braking index $n\ne3$, but retain the
angular dependence.

Following Section~6.1.1 of \citet{jones2001}, let $\Theta$ be the polar angles
made by the magnetic dipole with respect to the $z$-axis in the inertial frame
and $\Phi$ be the azimuthal angle with respect to the $x-z$ axes. Then our
generalisation of the vacuum point-dipole spin-down torque can be written as
\begin{align}
\ddot{\Phi} = -k \dot{\Phi}^{n} \sin^{2}\Theta,
\label{eqn: EM DE}
\end{align}
where $k$ is a positive constant. Rearranging \autoref{eqn: Theta} and
expanding about $\theta = 0$ up to $\mathcal{O}(\theta^{2})$, we find
\begin{align}
\begin{split}
\sin^{2}\Theta = & \sin^{2}\Theta_0
- 2\theta \sin\chi\cos\chi \sin\psi(t) \\
 & + \frac{1}{2}\theta^{2}\sin^{2}\chi\cos(2\psi(t)),
\end{split}
\label{eqn: sin 2 Theta}
\end{align}
where we have defined
\begin{align}
\sin^{2}\Theta_0 \equiv
\sin^{2}\chi + \theta^{2}\left(\cos^{2}\chi - \frac{\sin^{2}\chi}{2}\right),
\label{eqn: sin2Theta0}
\end{align}
a constant, while the second two
terms in \autoref{eqn: sin 2 Theta} provide the first and second harmonic
modulations in $\sin^{2}\Theta$.

In order to find approximate solutions to \autoref{eqn: EM DE}, we begin by
substituting the $\sin^{2}\Theta$ in \autoref{eqn: EM DE} with the time-averaged
constant $\sin^{2}\Theta_0$ value and solve to get
\begin{align}
\dot{\Phi}(t) = \dot{\Phi}_0\left[1 + (n-1)\frac{t}{\tauAge}\right]^{\frac{-1}{n-1}},
\label{eqn: zeroth order soln complete}
\end{align}
where
\begin{align}
\tauAge = \frac{|\dot{\Phi}_0|}{|\ddot{\Phi}_0|}
\approx \frac{1}{k|\dot{\Phi}_0|^{(n-1)} \sin^{2}\Theta_0}.
\end{align}

Now, we substitute \autoref{eqn: zeroth order soln complete} back into
\autoref{eqn: EM DE} along with the expanded, but complete variation in
$\sin^{2}\Theta$. To simplify the result, we expand in $t/\tauAge \ll 1$ and
write the result in terms of the spin frequency and its derivatives as
\begin{align}
\dot{\nu}(t) = \dot{\nu}_0 + \ddot{\nu}_0 t
-\dot{\nu}_0 \theta \left[
2\cot\chi \sin\psi(t)
- \frac{1}{2}\theta\cos(2\psi(t))\right]
\label{eqn: spin-down rate}
\end{align}

In this derivation, we make no assumptions on how $\psi(t)$ evolves. However,
since we are interested in cases where $\tauP \ll \tauAge$ we will assume that
\begin{align}
\psi(t) = \dot{\psi} t + \psi_0.
\end{align}
Then, following Sec~3 of \citet{jones2001}, but retaining the dependence on
$\theta$, we can write this as
\begin{align}
\psi(t) = -2\pi\frac{t}{\tauP} + \psi_0.
\end{align}
where
\begin{align}
\tauP \equiv \frac{1}{\epsilonP \nu(t) \cos\theta}
\end{align}

It can be shown that deriving this expression, but making the assumption $\chi
\ll 1$ in \autoref{eqn: sin 2 Theta} and throughout (rather than $\theta \ll
1$) is equivalent to the transformation $\theta\leftrightarrow\chi$ in
\autoref{eqn: spin-down rate}. This symmetry was discussed by
\citet{arzamasskiy2015} and fundamentally results from the symmetry of $\theta$
and $\chi$ in \autoref{eqn: Theta}. Because the same symmetry also exists in
our beam-width model (\autoref{eqn: beam-width signal model}), the
large-$\chi$ solutions presented in this work, can equally be interpreted as
small-$\chi$ solutions by interchanging $\theta$ and $\chi$.

\section{Consistency of posterior estimates in the \modelEpsDot-model}
\label{app: consistency}
\begin{figure}
\centering
\includegraphics[width=0.5\textwidth]{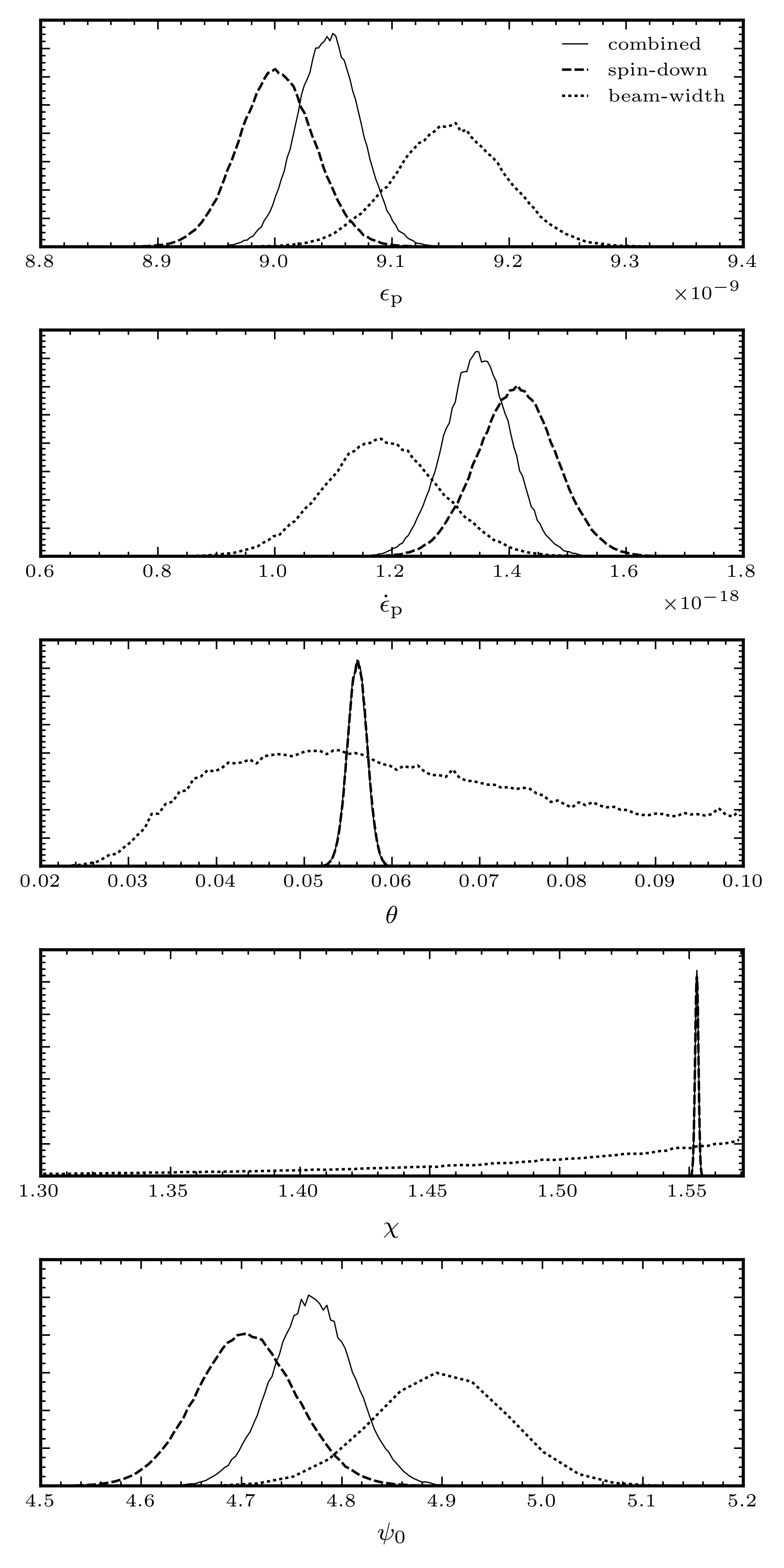}
\caption{Selected posterior distributions in the \modelEpsDot-model as
conditioned on the spin-down and beam-width data individually and the two
combined. Note that the $\theta$ and $\chi$ posteriors conditioned on the
beam-width data have been scaled by a factor of 10 so that they are visible
on the same scale as the strongly peaked spin-down and combined results.}
\label{fig: independent}
\end{figure}

For the base and \modelEpsDot-model, we investigated the behaviour when
conditioned on each data set (spin-down and beam-width) individually in
addition to the combined results presented in \autoref{sec: continuous
evolution of epsilon} and found that both data sets independently support the
\modelEpsDot-model over the base-model. In \autoref{fig: independent} we plot
the posteriors for the \modelEpsDot-model parameters that are common to both
the spin-down and beam-width parts of the model, excluding the frequency and
spin-down parameters which are dominated in all cases by the astrophysical
prior.

This figure demonstrates that the analysis performed on the two
individual data sets independently arrive at reasonably consistent posterior
distributions for these shared model parameters, with non-negligible overlap
between the posteriors.

For the two angles $\theta$ and $\chi$
the beam-width data does little to constrain the posteriors, with the results
even railing against the prior boundaries. Widening the prior (when
conditioning on the beam-width) solves this issue, but the posteriors remain
uninformative.  Comparing with the analysis of the combined data set, we see that
the combined posteriors are either a compromise of the individual
posteriors, when they are both informative, as is the case for $\epsilonP$, $\epsilondotP$, and $\psi_0$, or
they are dominated by the more informative spin-down data, as is the case for
$\theta$ and $\chi$.  As such, when using a combined data set, there is no
``tension'' (i.e. the two data sets preferring different solutions) and so their
log-odds sum approximately to the log-odds of the combined data set.

\bsp
\label{lastpage}
\end{document}